\documentstyle[12pt,epsfig]{article}

\catcode`\@=11
\def\@citex[#1]#2{%
\if@filesw \immediate \write \@auxout {\string \citation {#2}}\fi
\@tempcntb\m@ne \let\@h@ld\relax \def\@citea{}%
\@cite{%
  \@for \@citeb:=#2\do {%
    \@ifundefined {b@\@citeb}%
      {\@h@ld\@citea\@tempcntb\m@ne{\bf ?}%
      \@warning {Citation `\@citeb ' on page \thepage \space undefined}}%
      {\@tempcnta\@tempcntb \advance\@tempcnta\@ne%
      \@tempcntb\number\csname b@\@citeb \endcsname \relax%
      \ifnum\@tempcnta=\@tempcntb 
        \ifx\@h@ld\relax%
          \edef \@h@ld{\@citea\csname b@\@citeb\endcsname}%
        \else%
          \edef\@h@ld{\ifmmode{-}\else--\fi\csname b@\@citeb\endcsname}%
        \fi%
      \else
        \@h@ld\@citea\csname b@\@citeb \endcsname%
        \let\@h@ld\relax%
      \fi}%
    \def\@citea{,\penalty\@highpenalty\,}%
  }\@h@ld
}{#1}}

\def\@citeb#1#2{{[#1]\if@tempswa , #2\fi}}
\def\@citeu#1#2{{$^{#1}$\if@tempswa , #2\fi }}
\def\@citep#1#2{{#1\if@tempswa , #2\fi}}

\def\bcites{         
        \catcode`\@=11
        \let\@cite=\@citeb
        \catcode`\@=12
}

\def\upcites{         
        \catcode`\@=11
        \let\@cite=\@citeu
        \catcode`\@=12
}

\def\plaincites{      
        \catcode`\@=11
        \let\@cite=\@citep
        \catcode`\@=12
}

\newcount\hour
\newcount\minute
\newtoks\amorpm
\hour=\time\divide\hour by 60
\minute=\time{\multiply\hour by 60 \global\advance\minute by-\hour}
\edef\standardtime{{\ifnum\hour<12 \global\amorpm={am}%
        \else\global\amorpm={pm}\advance\hour by-12 \fi
        \ifnum\hour=0 \hour=12 \fi
        \number\hour:\ifnum\minute<10 0\fi\number\minute\the\amorpm}}
\edef\militarytime{\number\hour:\ifnum\minute<10 0\fi\number\minute}

\def\draftlabel#1{{\@bsphack\if@filesw {\let\thepage\relax
   \xdef\@gtempa{\write\@auxout{\string
      \newlabel{#1}{{\@currentlabel}{\thepage}}}}}\@gtempa
   \if@nobreak \ifvmode\nobreak\fi\fi\fi\@esphack}
        \gdef\@eqnlabel{#1}}
\def\@eqnlabel{}
\def\@vacuum{}
\def\marginnote#1{}
\def\draftmarginnote#1{\marginpar{\raggedright\scriptsize\tt#1}}
\def\draft{
        \pagestyle{plain}
        \overfullrule=2pt
        \oddsidemargin -.5truein
        \def\@oddhead{\sl \phantom{\today\quad\militarytime} \hfil
        \smash{\Large\sl DRAFT} \hfil \today\quad\militarytime}
        \let\@evenhead\@oddhead
        \let\label=\draftlabel
        \let\marginnote=\draftmarginnote
        \def\ps@empty{\let\@mkboth\@gobbletwo
        \def\@oddfoot{\hfil \smash{\Large\sl DRAFT} \hfil}
        \let\@evenfoot\@oddhead}
        \def\@eqnnum{(\theequation)\rlap{\kern\marginparsep\tt\@eqnlabel}%
        \global\let\@eqnlabel\@vacuum}  }

\def\nblack{            
        \def\ZZ{{Z \n{10} Z}}
        \def\NN{{N \n{14} N}}
        \def\CC{{C \n{11} C}}
        \def\RR{{R \n{11} R}}
        \def\QQ{{Q \n{12} Q}}
        \def\PP{{P \n{11} P}}
}

\def\eqalign#1{\null\,\vcenter{\openup\jot\m@th
  \ialign{\strut\hfil$\displaystyle{##}$&$\displaystyle{{}##}$\hfil
      \crcr#1\crcr}}\,}
\def\eqalignno#1{\displ@y \tabskip\centering
  \halign to\displaywidth{\hfil$\@lign\displaystyle{##}$\tabskip\z@skip
    &$\@lign\displaystyle{{}##}$\hfil\tabskip\centering
    &\llap{$\@lign##$}\tabskip\z@skip\crcr
    #1\crcr}}

\def\section{\@startsection {section}{1}{\z@}{3.ex plus 1ex minus
 .2ex}{2.ex plus .2ex}{\large\bf}}
\def\subsection{\@startsection{subsection}{2}{\z@}{2.75ex plus 1ex minus
 .2ex}{1.5ex plus .2ex}{\bf}}        

\def\appendix{{\newpage\section*{Appendix}}\let\appendix\section%
        {\setcounter{section}{0}
        \gdef\thesection{\Alph{section}}}\section}
\def\thefootnote{\arabic{footnote}}
\def\abstract{\if@twocolumn
\section*{Abstract}
\else 
\begin{center}
{\bf Abstract\vspace{-.5em}\vspace{0pt}}
\end{center}
\quotation
\fi}

\catcode`\@=12

\catcode`\@=11
\def\theequation{\arabic{equation}}
\catcode`\@=12

\nblack
\bcites

\nblack

\catcode`\@=11
\def\theequation{\thesection.\arabic{equation}}
\@addtoreset{equation}{section}
\@addtoreset{footnote}{section}
\@addtoreset{footnote}{subsection}
\catcode`\@=12

\newcommand{\beq}{\begin{equation}}
\newcommand{\bega}{\begin{array}}
\newcommand{\ea}{\end{array}}

\newcommand{\eeq}{\end{equation}}
\newcommand{\p}{\partial}
\newcommand{\U}{{\tilde U}}

\newcommand{\X}{{\tilde X}}

\begin{document}

\begin{titlepage}

\begin{center}
\hfill EFI-01-39 \\

\vskip 2.5 cm
{\large \bf Some remarks on D-branes in $AdS_3$}
\vskip 1 cm 
\renewcommand{\thefootnote}{\fnsymbol{footnote}}
{
Andrei Parnachev%
\footnote{
        andrei@theory.uchicago.edu
        }
and David A. Sahakyan%
\footnote{
        sahakian@theory.uchicago.edu
        }
}
\setcounter{footnote}{0}
\\
\vskip 0.5cm
{\sl Department of Physics, University of Chicago, \\

Chicago, IL 60637, USA\\ }

\end{center}

\vskip 0.5 cm
\begin{abstract}
We investigate the relation between the  algebraic
construction of  boundary states in $AdS_3$ and 
the target space analysis of D-branes and show the 
consistency of the two descriptions.
We compute, in the semiclassical regime, the overlap of a localized 
closed string state with  boundary states and identify
the latter with D-branes wrapping  conjugacy
classes in  $AdS_3$.
The string partition function on the disk is shown
to reproduce the spacetime DBI action. 
Other consistency checks are performed.
We also comment on the role of the
spectral flow symmetry of the underlying SL(2,R)/U(1) coset model
in constructing D-branes that correspond to degenerate
representations of SL(2,R).

\end{abstract}

\end{titlepage}

\section{Introduction}
\paragraph{}
Understanding D-branes in $AdS_3$ is an interesting
problem, which received much attention recently
\cite{GKS,BP,LOPT,PR,RR,HS,S,FOFS}.
The main difficulty stems from the non-compactness of the
space, which affects both the algebraic construction of
boundary states and the target space analysis.
The algebraic approach based on the modular transformation of
characters of the WZNW model \cite{Cardy} is a powerful technique
for identifying possible D-branes.
However, the application of this technique has been so far 
limited to rational CFTs.
In contrast, the SL(2,R) WZNW model, which describes bosonic
string theory on $AdS_3$, is not rational.
The alternative target space analysis of D-branes on
group manifolds has been developed quite
recently \cite{AS,ST,KaO,KlS}.
One of the important results of this analysis was the understanding that
maximally symmetric D-branes, which
preserve half of the current algebra, appear as surfaces
wrapping conjugacy classes in group manifolds.
In the SU(2) case, where  conjugacy classes are two-spheres,
this result was shown to
be consistent with the algebraic description \cite{MSM,FFFS}.
In fact, the spectrum of the theory on the brane,
obtained by expanding the DBI action to the second order
in fluctuating fields, was shown to exactly match the CFT result \cite{BDS}.
In the $AdS_3$ case important conjugacy classes are 
$dS_2$,  $H_2$ and $AdS_2$ surfaces. 
The target space analysis allows computation of the
DBI action for corresponding branes  \cite{BP}.
The algebraic description of D-branes in $AdS_3$
was pursued in \cite{GKS}, who used conformal bootstrap
in CFT on the disk (or the upper half plane)
to determine allowed boundary states.
More specifically, the crossing symmetry of the two-point function 
on the disk and the properties of degenerate operators in the SL(2,R)
WZNW model have been used to derive the constraint equation for 
the one-point function of an arbitrary primary of the current algebra.
By solving the constraint equation one may find all
possible boundary states.
In this note we try to understand these boundary states better.
Our main interest will be in their geometric interpretation,
but we also provide a review of conformal bootstrap on the disk
and re-derive some results of \cite{GKS} by different means.
\paragraph{}
This paper is organized in the following way.
Section 2 serves as a review of CFT on $AdS_3$ and  also summarizes
previous results on the geometry and the properties of 
extended D-branes in this background from the target space point of view.
This section also explains our notations.
Section 3 is devoted to the CFT analysis of D-branes.
Here we review  the conformal bootstrap of \cite{GKS} and
discuss D-branes that have a finite spectrum of open strings living 
on them.
It was shown in \cite{GKS} that such branes
correspond to finite dimensional representations of SL(2,R), which
are labeled by a discrete papameter.
We show that this result can be viewed as a
consequence of the spectral flow symmetry of the underlying SL(2,R)/U(1)
coset.
We next review the construction of boundary states that
have a continuous spectrum of open strings living on them.
These boundary states are labeled by a complex parameter.
When this parameter is pure imaginary (pure real),
the modular bootstrap provides a correspondence between
the D-brane and the principal continuous (principal discrete)
representation of SL(2,R).
In section 4, which is the main part of this paper, we
provide the geometric interpretation of boundary states that
possess a continuous spectrum of open strings.
By considering an overlap of a closed string state localized
in the target space with CFT boundary states, we recover  
D-branes that appear as surfaces of constant curvature
in Euclidean $AdS_3$\footnote{The Euclidean version of $AdS_3$
is the hyperbolic space, $H_3^+$.}.
We show that D-branes that correspond to principal
continuous  representations, wrap two-spheres in $H_3^+$.
We also argue that after analytic continuation to Lorenzian
$AdS_3$, the boundary states that correspond to principal continuous 
and principal discrete representations give rise to
D-branes that wrap $dS_2$ and $H_2$ conjugacy classes, respectively.
$H_2$ branes in $H_3^+$, which become surfaces wrapping
$AdS_2$ conjugacy classes in Lorenzian $AdS_3$, correspond 
to boundary states that are labeled 
by the parameter that is neither real nor imaginary.
We comment on the modular bootstrap interpretation of these states.
We provide support for the above picture by computing the string partition
sum on the disk, and comparing it with the DBI action of corresponding
D-branes.
Both quantities are divergent in the case of a D-brane that wraps an 
$H_2$ surface in $H_3^+$, since the area of the hyperbolic plane
is infinite.
We explain how this divergence can be regularized, and show that
once this is done, the CFT partition sum reproduces the DBI action
up to a finite normalization factor. 
In section 5 we discuss our results.
The solutions of the Knizhnik-Zamolodchikov
equations used in the main text appear in  Appendix A.
Appendix B studies the transformation of conformal blocks
under the action of the spectral flow.
Appendices C and D contain some useful formulae.

\section{Strings and branes in $AdS_3$}
The SL(2,R) group element can be written as
\begin{equation}
\label{matsl2r}
g=
\left( \begin{array}{ccc}
   X^0+X^1 &  X^2-X^3\\
   X^2+X^3 &  X^0-X^1
\end{array} \right),
\end{equation}
where
\begin{equation}
\label{hys}
(X^0)^2-(X^1)^2-(X^2)^2+(X^3)^2=1.
\end{equation}
One can think 
about SL(2,R) as a  hyperboloid defined by (\ref{hys}) in the 
four-dimensional space with the flat metric 
$ds^2=-(dX^0)^2+(dX^1)^2+(dX^2)^2-(dX^3)^2$.
One convenient parameterization of SL(2,R) is
\begin{equation}
   X^0+iX^3=\cosh \rho e^{it}, \qquad  X^1+iX^2=-\sinh \rho e^{-i\theta}.
\end{equation}
The range of coordinates is
$0 \le \rho \le \infty$, $0 \le \theta \le 2 \pi$, $0 \le t \le 2 \pi$,
and the metric is given by 
$ds^2=-\cosh^2 \rho \, dt^2+d\rho^2+\sinh^2 \rho \, d\theta^2$.
The unit radius $AdS_3$, which is the universal cover of SL(2,R)
is obtained by decompactifying the timelike direction $t$.
We can perform the Wick rotation $X^3=i\X^3$ 
which gives the hyperbolic space $H_3^+$.
This is a subspace of four-dimensional Minkowski space with 
the timelike coordinate $X^0$, parameterized by the
equation
\begin{equation}
\label{h3par}
(X^0)^2-(X^1)^2-(X^2)^2-(\X^3)^2=1, \qquad X^0 \ge 0.
\end{equation}
We will identify the point $g \in H_3^+$ with the 
complexified matrix (\ref{matsl2r}). 
A useful parameterization that we will need in the
following is
\begin{equation}
\label{matsl2r_2}
g=
\left( \begin{array}{ccc}
   \gamma \bar{\gamma} e^{\phi}+e^{-\phi} & -\gamma  e^{\phi}\\
   -\bar{\gamma}  e^{\phi} & e^{\phi}
\end{array} \right).
\label{param}
\end{equation}
$\gamma$ and $\bar{\gamma}$ are the coordinates on the
complex plane (the sphere) while $\phi$ is the radial coordinate.
The boundary of $H_3^+$ is at $\phi \rightarrow \infty$.
The metric takes the form $ds^2=d\phi^2+e^{2\phi} d\gamma d \bar{\gamma}$.
An important class of functions on  $H_3^+$ are those that transform 
as spin $j \equiv h-1$ representation 
of SL(2,R):
\begin{eqnarray}
\label{wf}
  {\tilde \Phi}_h(y,\bar{y}|g)=\frac{1-2 h}{\pi} \left[ \left(1 \quad  y \right) \;
  g \;
   \left(  \begin{array}{c}
   1 \\
   \bar{y}
   \end{array} \right) \right]^{-2h}=\frac{1-2 h}{\pi} \left( \frac{1}{
                      |\gamma-y|^2 e^{\phi}+e^{-\phi} } \right)^{2h}.
\end{eqnarray}
Here $y$ and $\bar{y}$ are the 
coordinates on the complex plane which parameterize the
function.
They also appear as the coordinates in the spacetime CFT
via AdS/CFT correspondence \cite{GKSg,KSg}\footnote{
   The spacetime coordinates are denoted by $y, \bar{y}$,
   not by $x,\bar{x}$ as in \cite{GKSg,KSg}. This choice of
   notation was made to avoid confusion with $X^i$ which
   parameterize $AdS_3$.}.
The introduction of  $y, \bar{y}$ \cite{FZ} is very convenient from the
technical point of view.
In particular,  Knizhnik-Zamolodchikov 
equations for four-point functions
containing degenerate operators become differential
equations that can be solved to obtain structure constants \cite{T}
or boundary states \cite{GKS}.
\paragraph{}
Bosonic string theory on $AdS_3$ is described by the  SL(2,R) WZNW
model.
It will be convenient to consider instead its Euclidean counterpart,
the $H_3^+$ WZNW model.
All our CFT results are therefore pertinent to the Euclidean
case.
The analytic continuation to Lorenzian $AdS_3$ may be performed
by Wick rotating one of the three spacelike coordinates in (\ref{h3par}).
The wavefunctions (\ref{wf}) present the semiclassical
expressions for operators in the   $H_3^+$ WZNW model.
The OPEs of the SL(2,R) currents with each other encode the
Kac-Moody algebra with central charge\footnote{The SL(2,R) WZNW model with 
central charge $k$ describes bosonic string theory in $AdS_3$ of radius $\sqrt{k}$.
In our geometric description of D-branes the coordinates are rescaled
so that the resulting space is $AdS_3$ of unit radius.} $k$.
\begin{eqnarray}
J^3(z) J^\pm(w) &\sim& \displaystyle  {\pm J^\pm(w)\over z-w}, \nonumber \\
J^3(z) J^3(w)  &\sim&   \displaystyle -{{k\over2}\over (z-w)^2}, \\
J^-(z)J^+(w)  &\sim&  \displaystyle  {k\over (z-w)^2}+{2J^3(w)\over z-w}. \nonumber 
\end{eqnarray}
These OPEs imply that the current modes defined by  an expansion
\beq
J^a(z)=\sum_{n=-\infty}^\infty {J^a_n\over z^{n+1}}
\eeq
satisfy the following commutation relations
\begin{eqnarray}
\left[J_n^3,J_m^3\right]&=&-\frac{k}{2} n \delta_{n+m,0}, \nonumber \\
\label{alg}
\left[J_n^3,J_m^{\pm}\right]&=&\pm J_{n+m}^{\pm}, \\
\left[J_n^+,J_m^-\right]&=&-2 J_{n+m}^3+ k n \nonumber \delta_{n+m,0}.
\end{eqnarray}
The functions (\ref{wf}) are promoted to  CFT
operators $\Phi_h(y,{\bar y};w,{\bar w})$ which are primaries of
the current algebra:
\begin{eqnarray}
J^3(z) \Phi_h(y,\bar y;w,\bar w) &\sim&
          \displaystyle  -{(y\partial_y+h)\Phi_h(y,\bar y;w,\bar w)
\over z-w}, \nonumber \\
\label{curope}
J^+(z) \Phi_h(y,\bar y;w,\bar w) &\sim&  \displaystyle
        -{\left(y^2\partial_y+2hy\right) \Phi_h(y,\bar y;w,\bar w)\over z-w}, \\
J^-(z) \Phi_h(y,\bar y;w,\bar w)
&\sim&  \displaystyle  -{\partial_y\Phi_h(y,\bar y;w,\bar w)\over z-w}. \nonumber
\end{eqnarray}
where  $w, \bar w$ stand for the coordinates on the
worldsheet.
The stress-energy tensor follows from Sugawara construction
\begin{equation}
\label{sut}
  T=\frac{1}{k-2} [-(J^3)^2+J^+J^-].
\end{equation}
The OPEs above imply that the global part of 
Kac-Moody symmetry can be interpreted as  global conformal
symmetry in spacetime.
In fact, the correspondence extends to the full infinite-dimensional
$\widehat{SL}(2,R)_k$ symmetry \cite{KSg}.
The operator  $\Phi_h(y,\bar{y};w, \bar w)$ is a primary under
both the spacetime and the  worldsheet conformal transformations
with scaling dimensions $h$ and 
\beq
\Delta_h=-u h (h-1)
\eeq
respectively. In the equation above we defined $u$ as
\beq
  u=\frac{1}{k-2}.
\eeq
Similar formulae hold in the antiholomorphic sector.
\paragraph{}
An important ingredient of the $H_3^+$ WZNW model is an operator
$\Phi_{-\frac{1}{2}}(y,\bar{y})$ which has a simple
semiclassical form
\beq
 \Phi_{-\frac{1}{2}}(y,\bar{y})=\frac{2}{\pi} 
         \left( |\gamma-y|^2e^{\phi}+e^{-\phi} \right).
\eeq
It is usually assumed that $\partial_y^2 \Phi_{-\frac{1}{2}}(y,\bar{y})=0$
holds as an operator equation.
This implies that the operator $\Phi_{-\frac{1}{2}}$
is degenerate
\beq
\label{ope3}
 [\Phi_{-\frac{1}{2}}] \, [\Phi_h] \sim
      [\Phi_{h+\frac{1}{2}}]+[\Phi_{h-\frac{1}{2}}].
\eeq
That is, the OPE 
of $\Phi_{-\frac{1}{2}}$ with a generic primary $\Phi_h$ contains only the
current algebra blocks of $\Phi_{h+\frac{1}{2}}$ and $\Phi_{h-\frac{1}{2}}$.
\paragraph{}
Let us now briefly review the results of Ref. \cite{BP} which
studies D-branes in $AdS_3$ background from the target space
point of view.
Depending on the gluing conditions for the holomorphic and antiholomorphic
currents at the boundary of the worldsheet, possible D-branes in
$AdS_3$ wrap conjugacy classes, twisted by a group automorphism
that descends from an algebra automorphism used in gluing
the currents.
Recall that the conjugacy class twisted by the automorphism
$w(h)$ is defined as
\begin{equation}
 {\cal W}_g^w= \{ w(h) g h^{-1}, \forall h \in SL(2,R) \, \}.
\end{equation}
In the case of the inner automorphism, $w(h)=g_0^{-1} h g_0$, $g_0 \in SL(2,R)$,
the set ${\cal W}_g^w$ is a left translation of the regular 
(untwisted) conjugacy class, so it is sufficient to consider
the latter.
In the parameterization (\ref{matsl2r}) the regular
conjugacy class is characterized by 
\begin{equation}
\label{ds2brane}
  \rm{tr} \, g=2 X^0=2 {\tilde C}.
\end{equation} 
It is a two-dimensional surface in the 
space (\ref{hys}) described by the following equation
\begin{equation}
(X^3)^2-(X^1)^2-(X^2)^2=1-{\tilde C}^2.
\end{equation}
Depending on the value of ${\tilde C}$, this surface can be a
hyperbolic $H_2^+$ plane ( ${\tilde C} < 1$), a lightcone
( ${\tilde C} = 1$) or a de-Sitter $dS_2$ plane ( ${\tilde C} > 1$). 
A coordinate system that will be useful for the description
of $dS_2$ D-branes is
\beq
X^0=\cosh {\tilde \psi}, \quad X^3=\sinh {\tilde \psi} \sinh {\tilde t}, \quad
X^1+iX^2=\sinh {\tilde \psi} \cosh {\tilde t} \, e^{i \phi}.
\eeq
(this coordinate system covers part of Lorenzian $AdS_3$).
The metric is given by
\beq
ds^2=d{\tilde \psi}^2+\sinh^2 {\tilde \psi}(-d {\tilde t}^2
                       +\cosh^2 {\tilde t} d \phi^2).
\eeq
After substitution of the solutions of equations of motion,
the DBI action of the D-brane located at constant
\beq
X^0= {\tilde C}=\cosh {\tilde \psi}_0
\eeq
is given by
\cite{BP}
\begin{equation}
\label{dbiac_ds2}
I_{dS_2}=i \sinh \tilde{\psi}_0 \, T_D \int d\tilde{t} 
d\phi \, \cosh \tilde{t}
\end{equation}
where $T_D$ is the (fixed) D-brane tension.
The action is imaginary because there is a supercritical electric field 
living on the $dS_2$ brane \cite{BP}.
$H_2$ D-branes with ${\tilde C}<1$ are best described in the coordinate system
\beq
 X^0=\cos {\tilde \tau}, \qquad X^3=\sin {\tilde \tau} \cosh \chi, \qquad 
  X^1+iX^2=\sin {\tilde \tau} \sinh \chi e^{i \phi} 
\eeq
where the metric is
\beq
 ds^2=-d {\tilde \tau}^2+\sin^2 {\tilde \tau}(d \chi^2+\sinh^2 \chi d \phi^2).
\eeq
In this coordinate system $H_2$ branes appear
as surfaces of constant
\beq
X^0={\tilde C}=\cos {\tilde \tau}_0.
\eeq
The DBI action of these branes is imaginary 
\beq
\label{dbiac_h2}
I_{H_2}= i T_D \sin {\tilde \tau}_0 \int d \chi d \phi \, \sinh \chi.
\eeq
Consider the fate of D-branes defined by (\ref{ds2brane}),
under the Wick rotation of $\X^3$ 
(of course, one may equivalently Wick rotate $X^1$ or $X^2$).
$X^0$ is now constrained as in (\ref{h3par}), and therefore
${\tilde C} < 1$ is ruled out.
$dS_2$ surfaces thus turn into two-spheres described by
\begin{equation}
  (\X^3)^2+(X^1)^2+(X^2)^2={\tilde C}^2-1,
\end{equation}
while the lightcone becomes a point (degenerate two-sphere).
The Euclidean action of such D-branes is given by (\ref{dbiac_ds2}).
No extra factors of $i$ appear, as $dS_2$  is
a timelike surface in $AdS_3$.
After the analytic continuation to $H_3^+$, the
coordinate $\tilde t$ takes a finite range $\tilde t \in [0, 2 \pi]$,
and therefore the DBI action, which contains
the volume of the D-brane, becomes finite.
\paragraph{}
Let us now look at conjugacy classes twisted
by the outer isomorphism
\beq
   w(h)=w_0^{-1} h w_0, 
   \qquad \omega_0=\left( \bega{ll} 0 & 1 \\ 1 & 0 \ea \right).
\eeq
These are $AdS_2$ surfaces defined by
\begin{eqnarray}
  \rm{tr} \, g=2 X^2&=&2  C, \\
(X^0)^2-(X^1)^2+(X^3)^2&=&1+C^2.
\end{eqnarray}
The convenient coordinate system
\beq
\label{ads_coord_system}
X^1=\cosh \psi \sinh w, \quad X^2=\sinh \psi, \quad
X^0+iX^3=\cosh \psi \cosh w e^{i t}
\eeq
has the metric
\beq
  ds^2=d \psi^2+ \cosh^2 \psi (-\cosh^2 w d t^2+dw^2).
\eeq
Note that the coordinate $t$ has an infinite range in $AdS_3$.
The DBI action of the $AdS_2$ D-brane located at constant
\beq
   X^2=C=\sinh \psi_0
\eeq
is now real \cite{BP}
\beq
\label{dbiac_ads2}
I_{AdS_2}= \cosh \psi_0 \, T_D \int d t dw \, \cosh w.
\eeq
The Euclidean counterparts of $AdS_2$ surfaces are $H^+_2$ planes
described by
\begin{equation}
(X^0)^2-(X^1)^2-(\X^3)^2=1+C^2.
\end{equation}
The Euclidean DBI action is still given by (\ref{dbiac_ads2}).
It contains an infinite volume of the hyperbolic plane.

\section{CFT boundary states}
\paragraph{}
In this section we will  review the conformal
bootstrap of \cite{GKS} and show how the crossing symmetry
of the two-point function 
$\langle  \Phi_{-\frac{1}{2}} \Phi_h \rangle$ on the disk leads to the
constraint on the one-point function of
the primary $\Phi_h$.
By solving this equation we obtain all allowed boundary states.
We first discuss D-branes that have a finite spectrum of
open strings living on them.
We show that the complementary constraint equation that
arises from the crossing symmetry of the ``dual'' two-point function 
$\langle \Phi_{\frac{k+1}{2}} \Phi_h \rangle$
implies that such D-branes are labeled by integers and correspond
to degenerate representations of SL(2,R).
Technically this follows from a simple relation between 
the original and the dual two-point functions, which is
a consequence of the spectral flow symmetry of the underlying
SL(2,R)/U(1) coset.
Hence, we re-derive the result, which was obtained in Ref. \cite{GKS} 
who used the two-point function 
$\langle  \Phi_{\frac{2-k}{2}} \Phi_h \rangle$ to derive
the complementary equation for the one-point function.
We next discuss boundary states that have a continuous spectrum
of open strings living on them, and therefore correspond to
extended D-branes in $AdS_3$.
This discussion will be important in the next section, where we turn 
to the geometric description of such D-branes.

\paragraph{}
We start by considering the holomorphic part of the
four-point function of  primary operators on the sphere.
The projective Ward identities for the worldsheet stress-energy
tensor and for the currents constrain the four-point function
 to have the following form
\begin{equation}\label{ward}
\langle \Phi_{h_0}(y_0,z_0) \Phi_{h_1}(y_1,z_1)  
           \Phi_{h_2}(y_2,z_2)  \Phi_{h_3}(y_3,z_3) \rangle=
   \prod y_{ij}^{\mu_{ij}} z_{ij}^{\nu_{ij}} 
   H\left(\bega{ll} h_0 &h_1\\ h_2 &h_3\ea,y,z\right),
\end{equation}
where $y_{ij} \equiv y_i-y_j$,  
$y$ is the projective invariant
\begin{equation}
y=\frac{(y_0-y_1)(y_2-y_3)}{(y_0-y_3)(y_2-y_1)},
\label{cross}
\end{equation}
and the non-zero $\mu_{ij}$ are 
\begin{eqnarray}
\mu_{03}&=&-2 h_0, \\
\mu_{31}&=&-h_1-h_3+h_0+h_2, \\
\mu_{32}&=&-h_2-h_3+h_0+h_1, \\
\mu_{21}&=&-h_0-h_1-h_2+h_3. 
\end{eqnarray}
The worldsheet variables $z_{ij}$, $z$, and $\nu_{ij}$  are 
defined similarly, with the
substitution $h_i \rightarrow \Delta_{h_i}$.
The function $H\left(\bega{ll} h_0 &h_1\\ h_2 &h_3\ea,y,z\right)$
satisfies the Knizhnik-Zamolodchikov equation.
As usual, it can be derived from (\ref{sut})
and the current Ward identities,
using the OPEs of the SL(2,R) currents with the primaries of
the Kac-Moody algebra.
The fact that the currents generate conformal symmetry in 
spacetime means that the equation can be conveniently 
written in terms of differential operators \cite{FZ}.
In our case the equation takes the form
\begin{eqnarray}
\label{kzeq}
&&\Bigg[ -z (z-1) (k-2) \partial_z
    +y (y-1) (z-y) \partial_y^2 \\ \nonumber
&&   \quad  +\left[ (\Delta{+}1) (-y^2{+}2 z y{-}z){-}2 h_0 y (y{-}1)
           {-}2 y h_1 (z{-}1) {-}2 (y{-}1) h_2 z \right] \partial_y
          \\ \nonumber
&&   \quad  {+}\left[2 h_0 \Delta (z{-}y) - 2 h_0 h_1 (z{-}1){-}2 h_0 h_2 z \right]
\Bigg]  H\left(\bega{ll} h_0 &h_1\\ h_2 &h_3\ea,y,z\right)=0,
\end{eqnarray}
where we introduced $\Delta=h_0+h_1+h_2-h_3$.
To define the theory on the disk\footnote{All our calculations
are performed on the upper half plane, which can be conformally mapped
to the disk.},
one must specify the gluing conditions for the
holomorphic and antiholomorphic currents at the boundary of the worldsheet.
Most of the subsequent discussion will be restricted to
the diagonal gluing, which implies the following form of the
one-point function
\begin{equation}
\label{1ptfun_lo}
   \langle \Phi_h(y,\bar{y}; z, \bar z) \rangle=
     \frac{U (h)}{\left(y-\bar{y}\right)^{2 h}
                        |z-\bar z|^{2 \Delta_h}   }=
              \frac{\U (h)}{|y-\bar{y}|^{2 h}
                        |z-\bar z|^{2 \Delta_h}   }.
\end{equation}
In the equation above we define $\U(h)$ and $U(h)$, which
are related as
\begin{eqnarray}
\label{fphase}
  \U(h)= i^{-2 h} U(h), \qquad {\rm Im} \, y>0; \\ \nonumber
  \U(h)=i^{2 h} U(h), \qquad {\rm Im} \, y<0.
\end{eqnarray} 
Note that because of (\ref{fphase})
it is impossible to have both $\U(h)$ and $U(h)$
to be completely $y$-independent.
We will see that while  $U(h)$ is $y$-independent, 
$\U(h)$ depends on the sign of $(y{-}\bar y)$ via (\ref{fphase}).
\paragraph{}
In the boundary CFT with the diagonal gluing, the two-point function on the 
disk, which contains  $\Phi_{-\frac{1}{2}}$,  can be written as
\beq
\langle \Phi_{-\frac{1}{2}}(y_1,z_1) \Phi_{h}(y_2,z_2)  \rangle=
\frac{|y_2-{\bar y_2}|^{-1-2h}}
                            {|y_1-{\bar y_2}|^{-2}} \;
                    \frac{|z_2-{\bar z_2}|^{-\frac{3 u}{2}-2\Delta_h}}
                            {|z_1-{\bar z_2}|^{-3 u}} 
  H\left(\bega{ll} -1/2 &h\\ -1/2 &h\ea,y,z\right),
\eeq
where $y$ and $z$ are the spacetime and worldsheet cross-ratios
\begin{equation}
  y= \frac{|y_1-y_2|^2}{|y_1-{\bar y_2}|^2} ;
  \qquad  z= \frac{|z_1-z_2|^2}{|z_1-{\bar z_2}|^2}.
\end{equation}
One can solve the Knizhnik-Zamolodchikov equation\footnote{See
Appendix A for the solutions
of the Knizhnik-Zamolodchikov equations in some special cases.}
and write the two-point function as
\begin{eqnarray}
\label{2corr}
 && \langle \Phi_{-\frac{1}{2}}(y_1,z_1) \Phi_{h}(y_2,z_2)  \rangle=
\frac{|y_2-{\bar y_2}|^{-1-2h}}
                            {|y_1-{\bar y_2}|^{-2}} \;
                    \frac{|z_2-{\bar z_2}|^{-\frac{3 u}{2}-2\Delta_h}}
                            {|z_1-{\bar z_2}|^{-3 u}} \\ \nonumber 
 && \qquad \times  \left[ C_+ \U(h+{1 \over 2}) \left( y H_1^+(z)+H_0^+(z) \right) 
            +C_- \U(h-{1 \over 2}) \left( y H_1^-(z)+H_0^-(z) \right) \right],
\end{eqnarray}
where $C_+$ and $C_-$ are the structure constants, and 
conformal blocks are given by
\begin{eqnarray}
\label{cblock1}
H_1^+&=&z^{u(1-h)} (1-z)^{\frac{3 u}{2}} F(u,1+2 u(1-h),1+u(1-2 h),z), \\
\label{cblock2}
H_0^+&=&{u 
        z^{1+u(1-h)} (1-z)^{\frac{3 u}{2}} \over (1+u(1-2h))}
         F(1+u,1+2 u(1-h),2+u(1-2 h),z), \\
\label{cblock3}
H_1^-&=& (2 h-1)^{-1} z^{uh} (1-z)^{\frac{3 u}{2}} F(1+u,2h u,1+u(2 h-1),z), \\
\label{cblock4}
H_0^-&=&  z^{uh} (1-z)^{\frac{3 u}{2}} F(u,2h u,u(2 h-1),z).
\end{eqnarray}
Here and elsewhere $F(A,B,C,z)$ is a hypergeometric 
function $_2F_1(A,B,C,z)$.
To fix the coefficients of conformal blocks one should
notice that the $|y_1-y_2|$ and $|z_1-z_2|$ dependence
of the $H_1^+$ and $H_0^-$ corresponds to the contribution
of the operators $\Phi_{h+{1 \over 2}}$ and $\Phi_{h-{1 \over 2}}$,
respectively.
The two-point function on the disk enjoys the crossing
symmetry.
In Appendix A we show that in addition to (\ref{2corr}),
it can also be written as
\begin{eqnarray}
\label{fusion_id}
&&\langle \Phi_{-\frac{1}{2}}(y_1,z_1) \Phi_{h}(y_2,z_2)  \rangle
= \frac{|y_2-{\bar y_2}|^{-1-2h}}
                            {|y_1-{\bar y_2}|^{-2}} \;
                    \frac{|z_2-{\bar z_2}|^{-\frac{3 u}{2}-2\Delta_h}}
                            {|z_1-{\bar z_2}|^{-3 u}} \\ \nonumber
 && \quad \times \left[ 
     B^+ (1-z)^{\frac{3 u}{2}} z^{u(1-h)} F(u,1+2 u(1-h),1+2u,1-z) (1-y)
+\cdots \right],
\end{eqnarray}
where we only displayed the conformal block that corresponds
to the contribution of the identity operator and its descendants
arising from the fusion of the operators $\Phi_h$ and $\Phi_{-{1 \over 2}}$ to
the boundary of the worldsheet.

\subsection{D-branes labeled by a discrete parameter}
\paragraph{}
Let us first discuss D-branes that  have a finite spectrum 
of  open strings living on them.
It will be convenient to assume that the one-point function
is normalized, i.e. is divided by the partition sum on the disk.
The conformal block that appears in (\ref{fusion_id})
has a simple behavior when two bulk operators are
taken close to the boundary of the worldsheet and spacetime.
Namely, its asymptotic behavior corresponds to the 
fusion of bulk primaries to the identity operators at
the boundary.
The normalization of the one-point function implies that
the coefficient of the conformal block factorizes as \cite{GKS,ZZ}
\begin{equation}
\label{factorization}
  B^+=\U(h)\U(-\frac{1}{2}).
\end{equation}
From  the equality of (\ref{2corr}) and  (\ref{fusion_id}), and
the transformation properties of hypergeometric
functions\footnote{See Appendix D for some properties
of hypergeometric functions.}
one can derive a constraint equation for the one-point function.
To do this, it is sufficient to match the terms containing $y$
in  (\ref{2corr}) and  (\ref{fusion_id}).
With the help of (\ref{hgzzz}) it may be shown that $H_1^+$ and
$H_1^-$  contain terms whose worldsheet dependence is precisely
the same as that of the conformal block in (\ref{fusion_id}).
Equating the coefficients gives the following relation
\begin{equation}
\label{constraint1}
   \U(-\frac{1}{2}) \U(h)= \frac{\Gamma(-2u)}{\Gamma(-u)} 
     \left[
   C_- \U(h{-}\frac{1}{2}) \, \frac{\Gamma(u(2h{-}1))}{\Gamma(2u(h{-}1))}
     -C_+ \U(h{+}\frac{1}{2}) \frac{\Gamma(1{+}u(1{-}2h))}{\Gamma(1{-}2hu)}
 \right].
\end{equation}
The expressions for the structure constants in the convenient
normalization were found in  \cite{GK} (see also  \cite{T})
by the free field techniques.
\begin{eqnarray}
\label{structure_consts}
C_+&=&\frac{2}{\pi} {\cal R}\left(-\frac{1}{2}\right), \\ \nonumber
C_-&=&\frac{2}{\pi}  {\cal R}\left(-\frac{1}{2}\right)
 \frac{\Gamma(2 u (h-1)) \Gamma(1+u (1-2 h))}{
       \Gamma(1+2 u (1-h)) \Gamma(u(2 h-1))}.
\end{eqnarray}
Here ${\cal R}(h)$  is the reflection amplitude 
\beq
\label{rh}
{\cal R}(h)=\frac{\Gamma(1+u (1-2 h))}{\Gamma(1-u (1-2 h))},
\eeq
which appears as the quantum mechanical
correction to the reflection symmetry \cite{T}
\beq
\label{refsymm}
\Phi_h(y, \bar y; z, \bar z)={\cal R}(h) \frac{2h-1}{\pi} \int
     d^2y' \, |y-y'|^{-4 h} \Phi_{1-h}(y', \bar y'; z, \bar z).
\eeq
At this point it is worth noting that eq.
(\ref{constraint1}) is invariant under
\beq
\U(h) \rightarrow i^{\pm 4h} \U(h).
\eeq
This is simply a reflection of the freedom that still exists in
the description. 
Namely, one can choose
either $U(h)$ or $\U(h)$ (see (\ref{1ptfun_lo}) for the definition)
to be independent of the sign of $(y{-}\bar y)$.
[Note that one cannot choose both of them to be completely
$y$-independent because of (\ref{fphase})].
To fix this freedom we may use the reflection symmetry \cite{GKS}.
By taking the expectation value on the upper half plane of both sides of
(\ref{refsymm}) it is not hard to see that it is $U(h)$ 
that must be $y$-independent.
The condition of the reflection symmetry (\ref{refsymm}) takes the form
\beq
\label{refsymm2}
\frac{U(h)}{(y-\bar y)^{2 h}}={\cal R}(h) \frac{2h-1}{\pi} \int
     d^2y' \, |y-y'|^{-4 h} \, \frac{ U(1-h)}{(y'-\bar y')^{2-2 h}}.
\eeq
The integral in the RHS of the equation above
was computed in \cite{GKS}; 
the result is
\beq
\label{conduh}
U(h)=-{\cal R}(h) U(1-h).
\eeq
This equation must be taken with some caution, as it
contains some regularized divergences.
We will return to this subject later in the paper.
It will be convenient to write $U(h)$ as
\begin{equation}
\label{def_of_f}
 U(h)= f(h) \Gamma(1+u (1-2 h)),
\end{equation}
The eq. (\ref{conduh}) translates into
\beq
\label{fasymm}
  f(h)=-f(1-h).
\eeq
The constraint equation (\ref{constraint1}) for $U(h)$ takes the form
\beq
\label{constraint2}
\pi \Gamma(1-u) f(-{1 \over 2}) f(h)=f(h-{1 \over 2})+f(h+{1 \over 2}).
\eeq
To derive this, we substituted (\ref{structure_consts}), (\ref{rh})
and (\ref{def_of_f}) into (\ref{constraint1}).
The solution of (\ref{constraint2}) which respects (\ref{fasymm}) is
\beq
\label{solf}
  f(h)=-\frac{1}{\pi \Gamma(1-u)} \, {\sin [ u \pi (2h'-1) (2 h-1) ]\over
\sin[u\pi(2h^\prime -1)]}.
\eeq
We will see that this solution is also consistent with the
spectral flow symmetry of the theory.
At this point $h'$ is an arbitrary complex number.
It was shown  in \cite{GKS} that $h'$
takes discrete values
\beq
\label{quanth}
  2h'-1 \in Z.
\eeq
Below we are going to re-derive this result using the
action of the spectral flow symmetry on the two-point function.
We will also provide verification by a direct computation.
The reader who is not interested in the details of this computation
may skip to the beginning of the subsection 3.2.
\paragraph{} 
The observation that will be important in the following
is that $\Phi_{k+1 \over 2}$ is a degenerate operator
which has the following fusion rules
\beq
\label{ope2}
[\Phi_{k+1\over 2}][\Phi_h]\sim [\Phi_{{k\over 2}-h-{1\over
2}}]+[\Phi_{{k\over 2}-h+{1\over 2}}].
\eeq
One way to see that this is true is to notice that the
Knizhnik-Zamolodchikov equation (\ref{kzeq}) takes a simple
form with two conformal blocks if $h_3={k +1 \over 2}$.
This was noticed long ago in the SU(2) case \cite{FZ},
and we explain how this comes about in Appendix A.
Another proof, which we will give below, utilizes the existence of the spectral flow symmetry.
In order to show that $\Phi_{k+1\over 2}$ is degenerate and  to
find the fusion rules of $\Phi_{k+1\over 2}$ with other operators in the theory
consider the state
\beq\label{degthet}
|\theta\rangle=(J^{+}_{-1})^2|{k+1\over 2}\rangle^{hw},
\eeq
where $|{k+1\over 2}\rangle^{hw}$ defined as
\beq
|{k+1\over 2}\rangle^{hw}\equiv \Phi_{k+1\over 2}(0)|0\rangle
\eeq
is a highest weight state, i.e.
\begin{eqnarray}
J_n^a|{k+1\over 2}\rangle^{hw}=0\quad n\geq 1;\\
J_0^+|{k+1\over 2}\rangle^{hw}=0.
\end{eqnarray}
We will also need a definition of the lowest weight state
below: the state $|a\rangle^{lw}$ is a
lowest weight state if the following conditions are met
\begin{eqnarray}
J_n^a|a\rangle^{lw}=0\quad n\geq 1;\\
J_0^-|a\rangle^{lw}=0.
\end{eqnarray}
We claim that $|\theta\rangle$ is a null state. This statement
is actually related to the fact that
\beq\label{deghalf}
\p^2_y\Phi_{-1/2}(y)\simeq 0,
\eeq
which means that the operator in the left-hand side of this equation
can be set to zero in all correlation functions.
Indeed (\ref{deghalf}) can be rewritten as
\beq\label{deghalf1}
|\tilde\theta\rangle^{hw}=(J_0^-)^2|-1/2\rangle^{hw}\simeq 0\quad \rm or
\eeq
$$
|\tilde\theta\rangle^{lw}=(J_0^+)^2|-1/2\rangle^{lw}\simeq 0,
$$
where
\beq
|-1/2\rangle^{lw}\equiv\lim_{y\rightarrow\infty}y^{-1}
\Phi_{-1/2}(y)|0\rangle
\eeq
and $|\tilde\theta\rangle^{lw}$, $|\tilde\theta\rangle^{hw}$
are the lowest and the highest weight states respectively.
As it was shown in \cite{MO}, the
spectral flow by minus one unit maps the lowest weight state
$|h\rangle^{lw}$  into the highest weight state
$|{k\over 2}-h\rangle^{hw}$ and also $J_0^+$ into $J_{-1}^+$.
The
spectral flow will map $|\tilde\theta\rangle^{lw}$ into the
highest weight state which is actually equal to $|\theta\rangle$.
We arrive to the conclusion that $|\theta\rangle$,
while being a descendant of the current algebra, is also a primary.
In a unitary theory this statement would mean that the operator
corresponding to $|\theta\rangle$ is null and can be set to zero in all
correlation functions. One should be aware of the fact
that in the context of the $H_3^+$ WZNW model this is just an
assumption, which seems to be a part of the definition of the theory.
\paragraph{}
One can also show directly that $|\theta\rangle$ is a primary, i.e.
\beq\label{prime}
\bega{l}
J^a_{n}|\theta\rangle=0;\quad {\rm for}\quad n>0\\
J^+_0|\theta\rangle=0.
\ea
\eeq
It is easy to see using (\ref{alg}) that the only non-trivial checks that
should be performed are the ones involving
$J^-_n$ and $J^3_n$ with $n=1,2$.
Indeed let $n>2$ and ${f^{ab}}_c$ be the structure constants of $sl(2,R)$ Lie algebra,
then
\beq
J^a_n(J^+_{-1})^2|{k+1\over 2}\rangle^{hw}=[J^a_n,(J^+_{-1})^2]|{k+1\over
2}\rangle^{hw}=\sum _b {f^{a+}}_b[J^b_{n-1},J^+_{-1}]|{k+1\over 2}\rangle^{hw}=
\eeq
$$
\sum_{b,c}{f^{a+}}_b {f^{b+}}_c J^c_{n-2}|{k+1\over 2}\rangle^{hw}=0.
$$
Using (\ref{alg}) we can
verify that  (\ref{prime}) holds.
Let us show 
that $J^-_1|\theta\rangle=0$
\beq
J_1^{-}(J_{-1}^+)^2|{k\over 2}+{1\over
2}\rangle^{hw}=((2J_0^3+k)J_{-1}^++J_{-1}^+(2J_0^3+k))|{k\over 2}+{1\over
2}\rangle^{hw}=0.
\eeq
The rest of the checks can be performed in a similar manner.
So we conclude that under our assumption the  operator corresponding
to $|\theta\rangle$
is null  and can be set to zero in all correlators.
Now this result can be used to derive the OPEs of the operator
$\Phi_{k+1\over 2}$ with other operators in the theory.
Suppressing the worldsheet dependence and the antiholomorphic part,
we have
\beq\label{ope}
\Phi_h(y)\Phi_{k+1\over 2}(0)=\sum_{h^\prime}
C^{h^\prime}_{{k+1\over 2}\,h}
y^{-{k+1\over 2}-h+h^\prime}\Phi_{h^\prime}(0)+\dots.
\eeq
Applying $(J_{-1}^+)^2$ to the (\ref{ope}) and using the OPE (\ref{curope})
we obtain the following constraint on the structure constants
\beq
\left[y^2\p_y+2hy\right]\left[y^2\p_y+2hy\right]
C^{h^\prime}_{{k+1\over 2}\,h}
y^{-{k+1\over 2}-h+h^\prime}=0,
\eeq
which can be written as
\beq\label{ope1}
\left(-{k+1\over 2}+h+h^\prime\right)\left(-{k\over
2}+{1\over 2}+h+h^\prime\right)
C^{h^\prime}_{{k+1\over 2}\,h}=0.
\eeq
We see that (\ref{ope1}) implies (\ref{ope2}).
Below we will show that this form of fusion rules is actually
fixed by spectral flow and follows from the fusion rules  (\ref{ope3}) of the
operator $\Phi_{-{1\over 2}}$.
\paragraph{}
Consider the three-point function
\beq
\langle \Phi_{h_1}(y_1;z_1) \Phi_{h_2}(y_2;z_2) \Phi_{h_3}(y_3;z_3) \rangle=
D(h_1,h_2,h_3) \, \eta(h_1,h_2,h_3;y_1,y_2,y_3;z_1,z_2,z_3).
\eeq
In this expression $\eta(h_1,h_2,h_3;y_1,y_2,y_3;z_1,z_2,z_3)$
contains the worldsheet and spacetime dependence, determined
by the current and conformal Ward identities (see \cite{T} for more details).
The explicit form of $D(h_1,h_2,h_3)$ is \cite{T}
\beq
\label{sc_rel0}
  D(h_1,h_2,h_3)=\frac{k-2}{2 \pi^3} \, 
   \frac{G({-}h_1{-}h_2{-}h_3{+}1) G({-}h_1{-}h_2{+}h_3) 
          G({-}h_1{-}h_3{+}h_2) G({-}h_2{-}h_3{+}h_1)}{
      G(-1) G(1-2 h_1) G(1-2 h_2) G(1-2 h_3)},
\eeq
where $G$ is some special function which satisfies
\begin{eqnarray}
\label{propg}
  G(x)&=&G(-x-k+1),\\ \nonumber 
  G(x-1)&=&\frac{\Gamma(1+{x \over k-2})}{\Gamma(-{x \over k-2})} \, G(x).
\end{eqnarray}
In eq. (\ref{sc_rel0}) the normalization of \cite{GK} is implied (the original
formulae of \cite{T} contain certain prefactors that are set to
unity in this normalization).
Using (\ref{propg}) one can show that
\begin{eqnarray}
  \frac{D({\hat h}_1,{\hat h}_2, h_3)}{
   D(h_1,h_2,h_3)}&=&
       \, \frac{ \Gamma(1+u[1-2 {\hat h}_1]) \Gamma(1+u [1-2 {\hat h}_2]) }{
           \Gamma(1+u[1-2 h_1]) \Gamma(1+u [1-2 h_2])}, \\ \nonumber
    {\hat h} &\equiv& {k \over 2} -h.
\end{eqnarray}
In Appendix B we show that this equation, 
which may also be written as
\beq\label{structur}
D(h_1,h_2,h_3)=
\sqrt{\langle\Phi_{h_1}\Phi_{h_1}\rangle
\langle\Phi_{h_3}\Phi_{h_3}\rangle
\over \langle\Phi_{{k\over 2}-h_1}\Phi_{{k\over 2}-h_1}\rangle
\langle\Phi_{{k\over 2}-h_3}\Phi_{{k\over 2}-h_3}\rangle}
D({k\over 2}-h_1,h_2,{k\over 2}-h_3),
\eeq
follows from the spectral flow symmetry.
(In the equation above  $\langle\Phi_h\Phi_h\rangle$ is a two point function of
the operator $\Phi_h$ stripped
of the spacetime and worldsheet dependence). Note that
$\langle\Phi_h\Phi_h\rangle$ is divergent due to the contribution of
zero modes but the divergences will cancel out in (\ref{structur}).
One can convince oneself that the structure 
constant that appears in the OPE of the degenerate
operator $\Phi_{k+1 \over 2}$ is related to $D({k+1\over 2},h,h^\prime)$
in a simple way
\beq
C^{h^\prime}_{{k+1\over 2}\,h}={D({k+1\over 2},h,h^\prime)\over
\langle\Phi_{h^\prime}\Phi_{h^\prime}\rangle}.
\eeq
It is interesting to note that because of the divergence in
$\langle\Phi_{h^\prime}\Phi_{h^\prime}\rangle$, the
structure constant $C^{{k\over 2}-h^\prime}_{{k+1\over 2}\,h}$
can only be nonzero when  $D({k+1\over 2},h,h^\prime)$ is divergent.
Now using (\ref{structur}) we can obtain the following relation
\beq
C^{{k\over 2}-h^\prime}_{{k+1\over 2}\,h}=
\sqrt{\langle\Phi_{k+1\over 2}\Phi_{k+1\over 2}\rangle
\langle\Phi_{h^\prime}\Phi_{h^\prime}\rangle
\over \langle\Phi_{-{1\over 2}}\Phi_{-{1\over 2}}\rangle
\langle\Phi_{{k\over 2}-h^\prime}\Phi_{{k\over 2}-h^\prime}\rangle}
C^{h^\prime}_{-{1\over 2}\,h}.
\eeq
Since the expression under the square root is non-zero
and finite, $C^{{k\over 2}-h^\prime}_{{k+1\over 2}\,h}$
and $C^{h^\prime}_{-{1\over 2}\,h}$ are non zero at the same values
of $h$ and $h^\prime$.
Hence the fusion rules (\ref{ope2}) indeed follow from (\ref{ope3}).
\paragraph{}
Let us now consider the two-point function on the disk containing 
this degenerate operator.
This two-point function reads
\begin{eqnarray}
\label{2corr_2}
\langle \Phi_{h}(y_1,z_1)  \Phi_{\frac{k+1}{2}}(y_2,z_2)  \rangle&=&
|y_2{-}{\bar y_2}|^{-k-1+2h}  |y_1{-}{\bar y_2}|^{-4 h} \\ \nonumber && \; \;
                    |z_2{-}{\bar z_2}|^{2\Delta_h-2\Delta_{\frac{k+1}{2}} }
                            |z_1{-}{\bar z_2}|^{-4 \Delta_h} 
       H\left(\bega{ll} h &(k+1)/2\\ h &(k+1)/2\ea,y,z\right). 
\end{eqnarray}
The conformal block that appears in the equation above
is given by
\beq
  H\left(\bega{ll} h &(k+1)/2\\ h &(k+1)/2\ea,y,z\right)=
      (y-z)^{-2 h} H_0'(z)+(y-z)^{-2 h-1} H_1'(z),
\eeq
with $H_0'$ and $H_1'$ related to $H_0$ and $H_1$ as 
\beq
\bega{l}
H_0^\prime(z)=z^h(1-z)^{-2\Delta_h+2\Delta_{-1/2}}H_0(z),\\
H_1^\prime(z)=z^{h+1}(1-z)^{-2\Delta_h+2\Delta_{-1/2}}(H_0(z)+H_1(z)),
\ea
\eeq
where we omit possible normalization factors that are
independent of the worldsheet coordinates.
We explain how the equations above follow from
the spectral flow symmetry in Appendix B, and verify
them directly by solving the corresponding Knizhnik-Zamolodchikov
equation in Appendix A.
The two-point function (\ref{2corr_2}) then takes the form 
that is similar to (\ref{2corr})
\begin{eqnarray}
\label{2corr_2m}
\langle \Phi_{h}(y_1,z_1)  \Phi_{\frac{k+1}{2}}(y_2,z_2)  \rangle&{=}&
|y_2{-}{\bar y_2}|^{{-}k{-}1{+}2h}  |y_1{-}{\bar y_2}|^{-4 h}
|z_2{-}{\bar z_2}|^{2\Delta_h{-}2\Delta_{\frac{k+1}{2}} }
                            |z_1{-}{\bar z_2}|^{-4 \Delta_h}  \\ \nonumber && \Bigg[
  {\tilde C}_+' \U ( {k \over 2} {-}[h{+}{1 \over 2}] ) \left[
           (y{-}z)^{-2 h} H_0'^{+}(z){+}(y{-}z)^{-2 h-1} H_1'^+(z) \right]+
\\ \nonumber &&  \; \;  {\tilde C}_-' \U ( {k \over 2} {-}[h{-}{1 \over 2}] ) \left[
           (y{-}z)^{-2 h} H_0'^{-}(z){+}(y{-}z)^{-2 h-1} H_1'^-(z) \right]
\Bigg],
\end{eqnarray}
where ${\tilde C}_+'$ and ${\tilde C}_-'$ 
are the structure constants, and the conformal blocks
are given by
\begin{eqnarray}
\label{cblock1_dual}
H_0'^+&=&z^{h+u h} (1-z)^{2 u h (h-1)} F(u, 2 h u, u (2 h-1),z), \\
\label{cblock2_dual}
H_1'^+&=&\frac{2 h}{2 h-1} z^{1+h+u h} (1-z)^{1+2 u h (h-1)},
          F(1+u, 1+2 h u, 1+u (2 h-1),z), \\
\label{cblock3_dual}
H_0'^-&{=}& \frac{u \, z^{1{+}h{+}u (1{-}h)} (1{-}z)^{2 u h (h{-}1)}}{1{+}u
(1{-}2 h)} F(1{+}u,1{+}2 u(1{-}h), 2{+}u (1{-}2 h),z), \\
\label{cblock4_dual}
H_1'^-&=&  z^{1+h+u (1-h)} (1{-}z)^{1+2 u h (h-1)}
          F(1{+}u,1{+}2 u(1-h), 1{+}u (1{-}2 h),z). 
\end{eqnarray}
To obtain the dual equation for the one-point function we
can again use the crossing symmetry.
The two-point function (\ref{2corr_2}) can also be written as
\begin{eqnarray}
\label{2corr_2_cr}
&&\langle \Phi_{h}(y_1,z_1)  \Phi_{\frac{k+1}{2}}(y_2,z_2)  \rangle =
|y_2{-}{\bar y_2}|^{-k-1+2h}  |y_1{-}{\bar y_2}|^{-4 h} 
                     \\ \nonumber
      &&\qquad \times |z_2{-}{\bar z_2}|^{2\Delta_h-2\Delta_{\frac{k+1}{2}} }
                            |z_1{-}{\bar z_2}|^{-4 \Delta_h}
      H\left(\bega{ll} h &h\\ (k+1)/2 &(k+1)/2 \ea,y,z\right).  
\end{eqnarray}
The analog of (\ref{fusion_id}) in this case is
\begin{eqnarray}
\label{fusion_id_2}
&&\langle \Phi_{h}(y_1,z_1)  \Phi_{\frac{k+1}{2}}(y_2,z_2)  \rangle {=}
 |y_2{-}{\bar y_2}|^{-k-1+2h}  |y_1{-}{\bar y_2}|^{-4 h} 
                     \\ \nonumber
 &&  \qquad \qquad \times \left[ 
     B'^+ \left( [1{-}y]{-}[1{-}z] \right)^{-2h} (1{-}z)^{2 u h (h-1)}z^{h+uh}
                                F(u,2 u h, 1{+}2 u, 1{-}z)
+\cdots \right].
\end{eqnarray}
Now we can use the same technology that was employed
in the derivation of (\ref{constraint1}).
The only difference is that we now need to match 
terms containing $(y{-}z)$ in the expressions (\ref{2corr_2m})
and (\ref{fusion_id_2}).
It is also convenient to make a shift $h \rightarrow k/2-h$
in the resulting equation, which then takes the form
\begin{eqnarray}
\label{constraint3}
&&  (-)^{k-2 h+1}  \U(\frac{k+1}{2}) \U(\frac{k}{2}-h)= \\ \nonumber && \qquad
  \frac{\Gamma(-2u)}{\Gamma(-u)} 
     \left[
   C_-' \U(h{-}\frac{1}{2}) \, \frac{\Gamma(u(2h{-}1))}{\Gamma(2u(h{-}1))}
     -C_+' \U(h{+}\frac{1}{2}) \frac{\Gamma(1{+}u(1{-}2h))}{\Gamma(1{-}2hu)}
 \right],
\end{eqnarray}
where  $C_+'$ and $C_-'$ are the structure constants that appear in 
the OPE of $\Phi_{k+1 \over 2}$ with $\Phi_{\frac{k}{2}-h}$ 
\beq
  [\Phi_{k+1 \over 2}]\,[\Phi_{\frac{k}{2}-h}] \sim
     C_-'[\Phi_{h-\frac{1}{2}}]+C_+'[\Phi_{h+\frac{1}{2}}].
\eeq
To find the relation between $C'_\pm$ and  $C_\pm$
we may use (\ref{sc_rel0}), which gives
\begin{eqnarray}
\label{sc_rel}
  \frac{C_{\pm}'}{C_{\pm}}&=&
       \frac{ \Gamma(1+u[1-2 {\hat h}_1]) \Gamma(1+u [1-2 {\hat h}_2]) }{
           \Gamma(1+u[1-2 h_1]) \Gamma(1+u [1-2 h_2])}, \\ \nonumber
  h_1&=&-{1 \over 2}, \qquad h_2=h, \\ \nonumber
     {\hat h_i} &\equiv& {k \over 2}-h_i.
\end{eqnarray}
Substituting 
\beq
\U(h)=i^{\pm 2h } \Gamma(1+u [1-2 h]) \, f(h),
\eeq
which is equivalent to (\ref{def_of_f}),
into (\ref{constraint3}) and (\ref{constraint1}), and using
 (\ref{sc_rel}) one can infer that
\beq
f({k \over 2}-h)= \pm f(h).
\eeq
Applied to (\ref{solf}), this leads to the result (\ref{quanth}).

\subsection{Extended D-branes}
In section 2 we encountered D-branes that wrap 
conjugacy classes in $AdS_3$ and therefore are extended in the target space.
This implies that the spectrum of open strings
living on such D-branes is continuous.
It was proposed in \cite{GKS} that the factorization property  (\ref{factorization})
is no longer valid for such D-branes and one should follow the
lines of \cite{FZZ}.
That is, one should consider  {\it unnormalized} one-point
function $U(h)$, rather than the normalized one, which played a central
role in the description of D-branes in the previous
subsection.
Further, the coefficient $B^+$ in the boundary expansion
of the two-point function (\ref{fusion_id}) takes the form
\beq
\label{factorization_e}
   B^+=i A_0 \U(h),
\eeq
where $A_0$ is the fusion coefficient of the operator $\Phi_{-{1 \over 2}}$
to the identity operator on the boundary
\beq
\label{ope12}
 \Phi_{-{1 \over 2}}(y,\bar{y}; w,\bar{w})=A_0 \frac{y-\bar{y}}{
                     |w-\bar{w}|^{-\frac{3 u}{2}}}+{\cdots}.
\eeq
The identity  (\ref{factorization_e}) implies that the
one-point function is again of the form (\ref{def_of_f})
with $f(h)$ satisfying the equation
\beq
\label{ca0}
   C A_0 f(h)=f(h-{1\over 2})+f(h+{1 \over 2}),
\eeq
where $C$ is some real number, whose precise value will not be needed
in our discussion.
The solution of this equation that respects the reflection symmetry 
(\ref{fasymm}) is
\beq
\label{fsin}
  f(h) =A \sin \left[ \Theta (2 h-1) \right],
\eeq
where $A$ is some prefactor.
The value of $A$ may in principle
be fixed via perturbative computation of the one-point function $U(h)$.
The analogous computation in Liouville theory \cite{FZZ}
implies that $A$ is independent of the parameter that labels the D-brane.
This observation was used in Ref. \cite{GKS}, who argued that
this is also the case in the $H_3^+$ WZNW model.
We will see that this conjecture is necessary for the consistency
of the CFT and spacetime descriptions. 
Eq. (\ref{ca0}) implies that $\Theta$ and $A_0$ are related as
\beq
\label{a0}
    C A_0=2 \cos \Theta.
\eeq
It is interesting to note that although (\ref{fasymm}),
and hence (\ref{fsin})
were derived for D-branes that preserve the diagonal 
SL(2,R) symmetry, they are also valid in case of other gluing conditions
on the SL(2,R) currents at the boundary of the worldsheet.
The gluing conditions that
give rise to 
$\langle \Phi_h(y, \bar y) \rangle= \frac{U(h)}{(1+y \bar y)^{2 h}}$
will be of particular importance.
The analog of (\ref{refsymm2}) in this case is
\beq
\label{refsymm3}
\frac{U(h)}{(1+y \bar y)^{2 h}}={\cal R}(h) \frac{2h-1}{\pi} \int
     d^2y' \, |y-y'|^{-4 h} \, \frac{ U(1-h)}{(1+y' \bar y')^{2-2 h}}.
\eeq
We  show that it leads to (\ref{fasymm}) in Appendix C. 
\paragraph{}
We conclude this section by a remark about the parameter
$\Theta$ which appears in the one-point function (\ref{fsin}).
Let us consider the annulus partition function of an open string stretched
between an extended D-brane labeled by $\Theta$ and a fundamental
D-instanton (the basic brane with $2h'-1=-1$).
In the closed string channel this partition function can be
written as
\begin{eqnarray}
\label{mb}
  Z_{(1,\Theta)}=\langle 1| e^{-2 \pi T {\cal H} } |\Theta \rangle
 & \sim & \int_{C^+} dh \int d^2 y
           \frac{U(h)_1}{(1+y \bar y)^{2 h}} \,
          \frac{U(h)_{\Theta}}{(1+y \bar y)^{2-2 h}} {
    q_c^{-u (h-{1 \over 2})^2}\over
          \eta^3(q_c)}, \\ \nonumber     
   q_c&=&\exp(-2 \pi T), \\ \nonumber
   \eta(q_c)&=&q_c^{1 \over 24}\, \prod_{n=1}^\infty (1-q_c^n), \\ \nonumber   
C^+ &\equiv& {1 \over 2}+i R.
\end{eqnarray}
Here $T$ and ${\cal H}$ are the time and the Hamiltonian respectively,
and the equality follows from inserting the complete set
of closed string states into the matrix element.
The subscripts in the one-point functions indicate the boundary 
states.
We also assumed that both D-branes carry the same gluing conditions,
which correspond to the one-point function of the form 
\beq
\label{lalala}
\langle \Phi_h(y,\bar y) \rangle ={U(h) \over (1+y \bar y)^{2h}}.
\eeq
This behavior of the one-point function suggests that the corresponding
D-brane does not introduce a boundary to the Euclidean spacetime.
Further, with this choice of gluing conditions
one does not encounter divergences in the
derivation of eq. (\ref{fasymm}), which appear for other gluing conditions.
All of this seems to imply that the D-instantons are described by the one-point function of the
form (\ref{lalala}) with
$U(h)$ given by (\ref{solf}) and (\ref{def_of_f}).
\paragraph{}
Let us continue to analyze (\ref{mb}).
It is clear that no additional $h$-dependent factor comes from the
$d^2y$ integration.
Hence, (\ref{mb}) reduces to
\beq
\label{mb1}
 Z_{(1,\Theta)} \sim \int_0^\infty d\lambda \,
       \lambda \sinh[2  \Theta \lambda] \exp(-2 \pi T u \lambda^2)  
  \, \eta^{-3}(q_c),
\eeq
which is essentially what has been computed in \cite{GKS},
who argued that it admits an interpretation in the open string
channel as a character of the principal continuous (principal
discrete) representation of ${\widehat SL}(2,R)$ algebra
if $\Theta$ is imaginary (real).
In this case
\beq
\label{idtheta}
\Theta=\frac{\pi (2 h'-1)}{k-2}
\eeq
is related to the spin $j'=h'-1$ representation of SL(2,R) \cite{GKS}.
The situation is more complicated for boundary states which carry other
gluing conditions, as an $h$-dependent term 
will modify the modular bootstrap (\ref{mb}).
For example, an analog of (\ref{mb1}) for the 
D-brane that is described by the one-point function of the form
$\langle \Phi_h(y, \bar y) \rangle =\frac{U(h)}{(y- \bar y)^{2h}}$,
will contain the additional $h$-dependent factor that is
computed in Appendix C.
The annulus partition function takes the  form
\beq
\label{mb2}
 Z_{(1,\Theta)} \sim \int_0^\infty d\lambda \,
        \left( \cosh[(2 \Theta + \pi) \lambda]-
                              \cosh[(2 \Theta -\pi) \lambda] \right)
    \exp(-2 \pi T u \lambda^2)  \, \eta^{-3}(q_c).
\eeq
This expression is not easily interpreted in the open string channel.
We will comment on this result later in the paper.

\section{Geometric interpretation}
In this section we consider D-branes in $H_3^+$
that appear as surfaces of constant $X^0$ (two-spheres) and $\X^3$
(hyperbolic planes)\footnote{See section 2 for the discussion
of the geometry of $H_3^+$, $AdS_3$, and D-branes wrapped on
conjugacy classes.}.
By Wick rotating one of the spacelike coordinates in $H_3^+$,
one may turn two-spheres into $dS_2$ surfaces. 
The fate of a D-brane at $\X^3=const$ depends on 
whether the coordinate which is Wick rotated lies within
the worldvolume of the D-brane ($X^1$ or $X^2$) or not ($\X^3$).
In the first case the D-brane becomes the $AdS_2$ brane in $AdS_3$.
In the second case it is necessary to have an imaginary value
of $\X^3$ for analytic continuation to make sense.
Depending on the value of $|\X^3|$, the resulting surface
can be a $dS_2$ conjugacy class ($|\X^3|>1$), a light cone ($|\X^3|=1$),
or an $H_2$ plane ($|\X^3|<1$).
We do not consider D-branes located at $X^1=const$ or at $X^2=const$,
since these two coordinates are on equal footing with $\X^3$ in $H_3^+$.
\paragraph{}
What we will show is that all of D-branes mentioned
above correspond to  boundary states 
in  CFT on the disk \cite{GKS}, which were reviewed in the
previous section.
We will see that the string partition sum on the disk 
reproduces the spacetime DBI action of corresponding D-branes.
The technique that we will be using is based on the fact
that the location of the brane in $H_3^+$ may be inferred
from its boundary wavefunction\footnote{The
similar computation in SU(2) gives D-branes that correspond
to conjugacy classes which in this case are simply  two-spheres \cite{MSM}.
Recently similar techniques were used for studying
D-branes in $AdS_3$ \cite{RR}. }. 
In the computations below  excited states of the string
are neglected, so everything boils down to quantum mechanics.
The quantum mechanical boundary wavefunction $\langle g| h' \rangle$
is a coordinate representation of the boundary state $|h'\rangle$.
The coordinate states $|g \rangle$ are normalized to a delta-function
which is defined with respect to the SL(2,R) invariant measure on $H_3^+$
\beq
  \langle g|g'\rangle=\delta(g-g'); \qquad g,g' \in H_3^+.
\eeq
A useful basis for normalizable
functions in $H_3^+$ is provided by the wavefunctions defined in (\ref{wf}) 
\cite{T,G}
\beq
\label{wfqm}
 \langle h,y,\bar{y}|g \rangle={\tilde \Phi}_h(y, \bar y|g).
\eeq
The corresponding states satisfy the completeness relation: 
\begin{eqnarray}
\label{kre}
 &&\int_{C^+} dh \; \int d^2y \; 
       |h,y,\bar y \rangle \langle h,y, \bar y|={\mathbf 1}, \\ \nonumber
 &&C^+ \equiv \frac{1}{2}+i R.
\end{eqnarray}
Using (\ref{kre}), the boundary wavefunction may be written as
\begin{equation}
\label{sam}
  \langle g| h' \rangle=\int_{C^+} dh \; \int d^2y \; 
        {\tilde \Phi}_h(y, \bar y|g)  \langle h,y,\bar{y}|h' \rangle.
\end{equation}
It is natural to identify
\beq
\langle h,y,\bar{y}|h' \rangle=\langle \Phi_h(y,\bar{y}) \rangle_{h'}.
\eeq
where the expression in the right-hand side stands for the
one-point function of $\Phi_h(y,\bar{y})$, stripped of the dependence of worldsheet
coordinates.
The  $y, \bar{y}$-dependence is determined
by the gluing conditions for the holomorphic and
antiholomorphic currents.

\subsection{D-Branes at $X^0=const$}

Let us consider  the gluing that gives rise to the
following one-point function
\beq
\label{ds2_phi}
\langle \Phi_h(y,\bar{y}) \rangle_{h'}=
        \frac{U(h)_{h'}}{(1+y\bar{y})^{2h}},
\eeq
where
\beq 
\label{1pt_ds2}
U(h)_{h'}=A \sin \left[ \frac{\pi (2 h'-1) (2 h-1)}{k-2} \right]
                   \Gamma (1+\frac{1-2 h}{k-2} ).
\eeq
We used (\ref{def_of_f}), (\ref{fsin}) and (\ref{idtheta})
in writing the formula (\ref{1pt_ds2}) for the one-point function.
To determine the location of the D-brane, which is described by 
(\ref{ds2_phi})--(\ref{1pt_ds2}),  it is necessary to compute
the overlap (\ref{sam}).
Plugging in (\ref{1pt_ds2}), (\ref{ds2_phi}) and (\ref{wf})
we can write (\ref{sam}) as
\begin{equation}
  \langle g| h' \rangle=\int_{C^+} dh \; \frac{1-2 h}{\pi} \int d^2y \;
     \frac{U(h)}{(1+y\bar{y})^{2h}}
   \frac{1}{\left( X^0{+}X^1{+}(X^0{-}X^1)y \bar{y} {+}2 X^2 y_1{+}2 \X^3 y_2 
              \right)^{2(1-h)}}.
\label{sam2}
\end{equation}
Let us introduce  $R=\sqrt{(X^2)^2{+}(\X^3)^2}$.
The integral over the $y$-plane can be rewritten as
\begin{eqnarray}
\label{someint}
  && \int d^2y \;
    \frac{1}{(1+y\bar{y})^{2h}}
   \frac{1}{\left( X^0{+}X^1{+}(X^0{-}X^1)y \bar{y} {+}2 X^2 y_1{+}2 \X^3 y_2
              \right)^{2(1-h)}}= \\ \nonumber \qquad 
  &&   (X^0{-}X^1)^{2h} \int d^2y \;
      \left((X^0{-}X^1)^2{+}y_1^2{+}y_2^2\right)^{-2h} 
          \left(1{+}R^2{+}y_1^2{+}2 R y_1{+}y_2^2\right)^{2(h-1)},
\end{eqnarray}
where we defined $y=y_1+iy_2$. Using the identity
\begin{equation}
\label{eden}
\int_0^\infty dt \, t^{\beta} e^{- \alpha t}=\alpha^{-\beta-1} \Gamma(\beta+1),
\end{equation}
we can rewrite (\ref{someint}) as
\begin{eqnarray}
 && \frac{(X^0{-}X^1)^{2h}}{\Gamma(2h) \Gamma(2-2h)}
 \int d^2y \, \int_0^\infty dt ds \, t^{2 h-1} s^{1-2 h} \\ \nonumber &&
 \qquad \qquad \qquad
      \exp \left( -t [(X^0{-}X^1)^2{+}y_1^2{+}y_2^2]
           -s [1{+}R^2{+}2 R y_1{+}y_1^2{+}y_2^2] \right).
\end{eqnarray}
Completing the square and integrating over $d^2y$ gives
\begin{equation}
\frac{(X^0-X^1)^{2h} \, \pi}{\Gamma(2h) \Gamma(2-2h)} \int_0^\infty dtds \;
      \frac{t^{2h-1} s^{1-2h}}{t+s}
      \exp\left( \frac{R^2 s^2}{t+s}-t(X^0-X^1)^2-s(1+R^2) \right).
\end{equation}
Writing $t=\alpha s$ and integrating over $s$ we obtain
\begin{equation}
\frac{(X^0-X^1)^{2h} \pi}{\Gamma(2h) \Gamma(2-2h)} 
   \int_0^\infty d \alpha \frac{\alpha^{2h-1}}{1+
         \alpha (1+R^2)+(1+\alpha) \alpha (X^0-X^1)^2}.
\end{equation}
With the help of (\ref{h3par}) this can be simplified to
\begin{equation}
\label{almost_ti}
\frac{\pi}{\Gamma(2h) \Gamma(2-2h)} 
   \int_0^\infty d \alpha \frac{\alpha^{2h-1}}{1+2 \alpha X^0+\alpha^2}.
\end{equation}
Note that this integral depends only on $X^0$.
Recall that in $H_3^+$ $X^0>1$.
Writing 
\beq
X^0=\cosh \tilde{\psi},
\eeq
and using the identity \cite{GR}
\begin{eqnarray}
\label{ti_1}
\int_0^\infty \frac{x^{\mu-1} dx}{(x+\beta)(x+\gamma)}=
 \frac{1}{\gamma-\beta} (\beta^{\mu-1}-\gamma^{\mu-1}) \frac{\pi}{\sin(\pi \mu)} \\
\nonumber \quad |\arg \beta|, |\arg \gamma|< \pi, \quad 0< {\rm Re} \mu <2,
\end{eqnarray}
the integral (\ref{someint}) becomes
\beq
\label{iii1}
\frac{\pi}{2 h-1} \, \frac{\sinh[(2 h-1) {\tilde \psi}]}{\sinh {\tilde \psi}}.
\eeq
Substituting this into (\ref{sam2})
we obtain the following expression for the
boundary wavefunction
\begin{equation}
 \langle g |h' \rangle=-A  \int_{C^+} dh \; U(h) 
         \frac{\sinh[ (2h-1) {\tilde \psi}]}{\sinh {\tilde \psi}}.
\label{int_h}
\end{equation}
We may now substitute (\ref{1pt_ds2}) into this expression
and perform  the integration.
Before we do this, however, let us note
that although the integral formally runs over $h=\frac{1}{2}+i \lambda$,
$\lambda \in R^+$, our analysis is only justified in
the semiclassical limit, and therefore the terms of the order $h/(k-2)$
that appear in the argument of the Gamma function in (\ref{1pt_ds2})
should be neglected.
(This is similar to what happens in the SU(2) case \cite{MSM}.)
For the D-brane labeled by 
\beq
\label{pclabel}
h'=\frac{1}{2}+i \lambda',
\eeq
the boundary wavefunction becomes
\begin{equation}
 \langle g |h' \rangle=   - \frac{A}{\sinh \tilde{\psi}} \int_0^\infty
  d\lambda \; 
     \sin[\frac{2 \pi \lambda'  }{k-2} \lambda] 
        \sin[\tilde{\psi} \lambda].
\eeq
Up to an inessential numerical prefactor, this
is simply a delta-function
\beq
\langle g |h' \rangle \sim  A \, \frac{\delta(\tilde{\psi}-\frac{2 \pi \lambda'}{k-2})
          -\delta(\tilde{\psi}+\frac{2 \pi \lambda'}{k-2})
}{
                                         2 \sinh \tilde{\psi}}=
A \, \delta(X^0-\cosh {\tilde \psi}_0),
\label{loc2}
\end{equation}
where we defined
\beq
  {\tilde \psi}_0=\frac{2 \pi \lambda'}{k-2}.
\eeq
That is, the boundary state labeled by a parameter
from the principal continuous series (\ref{pclabel}), 
gives rise to the D-brane which appears as the surface
of constant
\beq 
X^0={\tilde C}=\cosh {\tilde \psi}_0 
\eeq
(equivalently, the two-sphere of radius $\sqrt{{\tilde C}^2-1}$)
in $H_3^+$.
The CFT partition function
\begin{equation}
\label{cftac_ds2}
U(h=0)_{h'} \sim i \sinh[\frac{2 \pi \lambda'}{k-2}]=i \sinh {\tilde \psi_0},
\end{equation} 
reproduces (\ref{dbiac_ds2}) up to a real normalization
constant.
\paragraph{}
Let us now look at the fate of the D-brane described by 
(\ref{pclabel}) under the analytic continuation $\X^3 \rightarrow iX^3$.
The two-sphere $X^0={\tilde C}$ becomes the $dS_2$ surface
in Lorenzian $AdS_3$.
The important difference from the Euclidean case is that
$X^0$ becomes an unrestricted coordinate.
In the computation of the matrix element above, we used the
positivity of $(X^0-X^1)$, which is equivalent to the positivity of $X^0$.
It is not hard to see that under the change of sign in $X^0$,
the integrand in (\ref{sam2}) picks up an $h$-dependent phase $(-)^{2h}$.
As was explained in the previous section, conformal bootstrap is
insensitive to such a phase.
Therefore, in  light of the above discussion, it seems natural to
propose that the multiplication of the one-point function by $(-)^{2h}$
corresponds to the reflection $X^0 \rightarrow -X^0$  in the target space.
In other words, the one-point functions that correspond to 
D-branes located at $X^0={\tilde C}$ and at
$X^0=-{\tilde C}$ differ by $(-)^{2h}$.
\paragraph{}
Before we proceed to other gluing conditions, let us comment
on the fate of D-branes defined by (\ref{ds2_phi}) as
above but with $h'$ not belonging to $C^+$.
For an illustration, let us take $h'=\frac{1}{2}+\mu'$ to
be real.
Then the boundary wavefunction (\ref{int_h}) turns into 
\beq
\label{ediv_int}
\langle g |h' \rangle  = -\frac{A}{\sinh \tilde{\psi}} \int_0^\infty
  d\lambda \; 
     \sinh[\frac{2 \pi \mu'  \lambda}{k-2}] 
        \sin[\tilde{\psi} \lambda].
\eeq
This is a divergent integral with the integrand being
an  oscillating function whose amplitude increases
exponentially as $\lambda \rightarrow \infty$.
The only sensible answer for this integral is zero.
One can see this directly by rewriting the integrand as a sum
of exponents and using analytic continuation.
This is also physically reasonable since, as we will see,
D-branes with real $h'$ appear as  $H_2$ surfaces
in Lorenzian $AdS_3$ and therefore should not appear in $H_3^+$.
\paragraph{}
It is interesting to observe that 
D-branes with  $h'$ real, prohibited in the Euclidean case,
become allowed after the analytic continuation 
to Lorenzian  $AdS_3$.
In the parameterization 
\beq
    X^0=\cos {\tilde \tau},
\eeq
the boundary wavefunction (\ref{int_h}) becomes
\begin{equation}
 \langle g |h' \rangle = -\frac{A}{\sin {\tilde \tau}} \int dh \; 
       \sin [ \frac{2 \pi \mu' (2 h-1)}{k-2} ]
     \sin[ (2h-1) {\tilde \tau}].
\label{int_h_h3}
\end{equation}
It is clear that it is necessary to rotate the contour of integration
to the real axis in order for this integral to make sense.
This is quite reasonable, since in Lorenzian $AdS_3$ 
the normalizable wavefunctions include the highest and lowest
weight states with real $h \ge 1/2$.
The integral over $h$ therefore 
corresponds to the trace over these states. 
Performing the integral gives
\beq
\langle g |h' \rangle  \sim 
 A\, \frac{\delta(\tilde{\tau}-\frac{2 \pi \mu'}{k-2})}{
                                         \sin \tilde{\tau}}.
\eeq
Hence we recover the $H_2^+$ brane 
located at $X^0=\cos[ \frac{2 \pi \mu'}{k-2}]=\cos {\tilde \tau}_0$.
The string partition sum
\beq
U(h=0)_{h'} \sim \sin {\tilde \tau}_0,
\eeq
reproduces the corresponding DBI action (\ref{dbiac_h2}).

\subsection{D-Branes at $X^3=const$}
Let us now explore the gluing conditions that 
give rise to the following one-point function
\begin{equation}
\label{ads2_phi}
   \langle \Phi_h(y,\bar{y}) \rangle_{\Theta}=
              \frac{U(h)_{\Theta}}{\left(y-\bar{y}\right)^{2 h}},
\end{equation}
where 
\beq
U(h)_{\Theta}=A \sin [ \Theta (2 h-1) ] \,
                   \Gamma (1+\frac{1-2 h}{k-2} ).
\eeq
In the equations above we introduced a complex parameter $\Theta$,
which labels the boundary state, in accord with the discussion at 
the end of the previous  section.
The boundary wavefunction now takes the form
\begin{equation}
  \langle g| \Theta \rangle=\int_{C^+} dh \; \frac{1-2 h}{\pi} \int d^2y \;
     \frac{U(h)}{(2 i y_2)^{2 h}}
   \frac{1}{\left( X^0{+}X^1{+}(X^0{-}X^1)y \bar{y} {+}2 X^2 y_1{+}2 \X^3 y_2 
              \right)^{2 (1-h)}}.
\label{sam_ads2}
\end{equation}
After integration over  $y_1$ the result depends only on $\X^3$:
\begin{equation}
   \langle g| \Theta \rangle=\frac{1-2 h}{\pi}
                    \frac{ \Gamma(\frac{3}{2}-2 h) \Gamma(\frac{1}{2})}{
                                \Gamma(2-2h)}
     \int_{C^+} dh \;  U(h) \int dy_2 \; 
               (2 i y_2)^{-2 h} \left[ (y_2+\X^3)^2+1 \right]^{-\frac{3}{2}+2 h}.
\label{ghi}
\end{equation}
Using (\ref{eden}), completing the square and performing the integral over
$y_1$ we obtain
\begin{eqnarray}
&& \frac{ \Gamma(\frac{3}{2}-2 h) \Gamma(\frac{1}{2})}{
                                \Gamma(2-2h)} \, \int dy_2 \;  (2 i y_2)^{-2 h} 
 \left[ (y_2+\X^3)^2+1 \right]^{-\frac{3}{2}+2 h}= \\ \nonumber
&&\qquad \frac{ \pi}{\Gamma(2 h) \Gamma(\frac{3}{2}-2 h)}
  \int_0^\infty dt ds \, t^{2h-1} s^{-2 h}
    \exp( -s +2 i \X^3 t-t^2/s).
\label{tsint}
\end{eqnarray}
Writing $t=\alpha s$ and  integrating over $s$ we obtain
\begin{equation}
\label{almost_ti_2}
   \frac{\pi}{\Gamma(2h) \Gamma(2-2h)} 
   \int_0^\infty d \alpha \frac{\alpha^{2h-1}}{1-2 i \alpha \X^3+\alpha^2}.
\end{equation}
which is very similar to (\ref{almost_ti}).
We performed some formal manipulations, so it would be
nice to have an independent check of the formula above.
Fortunately, there is a straightforward way of relating
(\ref{almost_ti_2}) to the computations we have done
previously.
Namely, we can set $\X^3=iX^3$ to be imaginary.
This converts  (\ref{almost_ti_2}) into (\ref{almost_ti}) with
the substitution $X^3 \rightarrow X^2$.
Therefore we indeed recover $dS_2$ and $H_2$ branes of
the previous subsection.
As in the case of $dS_2$ ($H_2$) branes at $X^0=const$,
the identification (\ref{idtheta}) relates the complex parameter
in the one-point function to the principal continuous
(discrete) representation of SL(2,R).
\paragraph{}
For real $\X^3$ let us introduce the parameterization
\beq
\X^3=\sinh \psi.
\eeq
The integral (\ref{almost_ti_2}) becomes
\begin{equation}
\frac{\pi}{\Gamma(2h) \Gamma(2-2h)} 
   \int_0^\infty d \alpha \frac{\alpha^{2h-1}}{
   (\alpha+e^{\psi-i \frac{\pi}{2}}) (\alpha+e^{-(\psi-i \frac{\pi}{2})})  }.
\end{equation}
Using (\ref{ti_1}) this can be written as
\beq
\label{iii2}
 \frac{\pi}{2 h-1} \, \frac{\sinh[ (2 h-1) (\psi-i{\pi \over 2})]}{
                              \cosh \psi},
\eeq
and therefore the boundary wavefunction becomes
\begin{equation}
\label{int_df}
 \langle g |\Theta \rangle =- \frac{A}{\cosh \psi} \int_0^\infty  d\lambda  \; 
     \sin[i \Theta  \lambda] 
        \sin[ \lambda (\psi - i {\pi \over 2})].
\end{equation}
When $\Theta$ is of the form
\beq
\label{strangeresult}
 \Theta=i \psi_0 +{\pi \over 2},
\eeq
the integral in (\ref{int_df}) produces the delta-function
\beq
   \langle g |\Theta \rangle \sim A\,\frac{\delta(\psi-\psi_0)}{
                     \cosh \psi_0}.
\eeq
Hence we recover the $H_2$ brane located at 
\beq
\X^3=C=\sinh \psi_0,
\eeq
which becomes the $AdS_2$ brane after the Wick rotation of 
$X^1$ or $X^2$.
Note that in this case the spacetime reflection $\X^3\rightarrow -\X^3$
corresponds to the complex conjugation of $\Theta$ in the worldsheet description.  
The string partition sum on the
disk
\beq
\label{sps_ads2}
 U(0)_{\Theta} \sim \sin[i \psi_0+{\pi \over 2}]=\cosh \psi_0,
\eeq
reproduces the DBI action  (\ref{dbiac_ads2}) up to some
{\it real} normalization factor.
\paragraph{}
To perform another consistency check of (\ref{strangeresult})
let us recall the construction of \cite{GKS}.
According to \cite{GKS}, various boundary states in the $H_3^+$
WZNW model on the disk correspond to the value of  boundary
perturbation constant $E$ which enters the boundary CFT action.
The latter can be written in the free-field representation as
\beq
{\cal S_B}=E \int dz \beta e^{- \phi(z)},
\eeq
where the integral runs over the real line, which is the boundary of
the upper half plane.
The free-field (Wakimoto) representation of the $H_3^+$ 
WZNW model contains the linear dilaton field $\phi$,
together with the ($\beta$, $\gamma$) system of conformal
weight (1,0).
The details of the Wakimoto representation will not be important
to us, and we only note that the value of 
the ($k$-dependent) slope of the linear
dilaton theory is consistent with the worldsheet and
spacetime behavior  of correlators demanded by Ward identities.
The OPE (\ref{ope12}) can be derived perturbatively,
by expanding the exponential of the boundary action to the
first order \cite{GKS}.
Using the free-field representation of the operator 
\beq
\Phi_{-{1 \over 2}}(y,\bar{y};w,\bar{w})=
   \frac{2}{\pi} {\cal R} (-1/2) \gamma(w)  e^{ \phi(w)} y+
  {\cdots}
\eeq
gives the following equation
\beq
\label{opepert}
  \frac{A_0 }{
                     |w-\bar{w}|^{-\frac{3 u}{2}} }=
  E \int dz \, \beta(z) e^{-\phi(z)}
    \; \frac{2}{\pi} {\cal R} (-1/2) \gamma(w)  e^{ \phi(w)}.
\eeq
Here ${\cal R} (-1/2)$ is a real reflection coefficient necessary
to account for the multiplicative renormalization of the operator
$\Phi_{-{1 \over 2}}$.
By using the free field OPEs, it is not hard to see
that the right hand side of (\ref{opepert}) is a pure
imaginary number.
Therefore $A_0$ must be pure imaginary.
Interestingly, the relation (\ref{a0}) 
implies that this can only be the case if $\Theta$ is
of the form (\ref{strangeresult}).
\subsection{Volume divergences}
The expressions (\ref{dbiac_ds2}) and (\ref{dbiac_ads2})
contain  divergent integrals over the worldvolume of  corresponding
conjugacy classes.
We have been so far a little bit cavalier matching
the CFT partition function on the disk with the spacetime
DBI action.
The problem is that the divergences need to be regularized,
and it is not {\it a priori} obvious that the regularization will not introduce
an extra dependence on the location of the brane, which
would spoil the relation between the CFT and the spacetime results
for the D-brane action.
The purpose of the discussion below is to show that this does not
happen.
\paragraph{}
In the Euclidean $AdS_3$ the D-brane labeled by the
principle continuous representation of SL(2,R) is a surface
of constant $X^0={\tilde C}$, which is a two-sphere
of radius $\sqrt{{\tilde C}^2-1}$.
The DBI action (\ref{dbiac_ads2}) becomes finite, as the timelike
coordinate takes a finite range ${\tilde t} \in [0,2 \pi]$.
The situation with the D-brane located at constant $\X^3=C$
is more tricky, as it is an infinitely stretched $H_2$
plane (in the Euclidean case), with the DBI action (\ref{dbiac_ads2}) containing
the divergent integral
\beq
\label{dint}
  \int dt \,dw\, \cosh w.
\eeq
It is clear that this divergence is a manifestation of
the fact that the $H_2$ plane has infinite volume.
Let us now understand how this is reflected
in the CFT analysis.
Recall that the CFT partition sum on the
disk is given by $U(h=0)$, which is related to $U(h=1)$
by the reflection symmetry.
In the case of a D-brane at $X^0={\tilde C}$, we can
use (\ref{refsymm3}) to write
\beq
\label{refsymm3d}
U(0)= -{1 \over \pi} \frac{\Gamma(1+u)}{\Gamma(1-u)} \int
     d^2y' \,  \frac{ U(1)}{(1+y' \bar y')^{2}}= 
    -\frac{\Gamma(1+u)}{\Gamma(1-u)} \, U(1).
\eeq
Hence, the partition sum $U(0)$ is indeed finite. 
Recall that the corresponding DBI action is also finite.
Hence the description is free of volume divergences,
and the CFT partition sum (\ref{cftac_ds2}) is 
equal to the DBI action (\ref{dbiac_ds2}) up to a finite constant.
Something interesting happens when we write the partition
sum for a D-brane located at $\X^3=C$:
\beq
\label{refsymm2d}
U(0)= -{1 \over \pi} \frac{\Gamma(1+u)}{\Gamma(1-u)} \int
     d^2y' \,  \frac{ U(1)}{(y'-\bar y')^{2}}.
\eeq
This expression contains  divergent integral, which, in light
of the discussion above, must therefore signify the volume divergence.
As we mentioned earlier, the result (\ref{fasymm})
may  contain regularized infinities for the gluing
conditions that are associated with D-branes located at $X^1,X^2,X^3=const$.
\paragraph{}
The divergent integral that appears in (\ref{refsymm2d})
must be equal to (\ref{dint})\footnote{One 
may introduce any suitable regularization, like
restricting the range of coordinates in $AdS_3$, to define the 
divergent integrals.}, up to a finite factor.
To see that this factor is independent of $C$, it is useful to
consider an overlap of the two boundary states  $|\Theta_1 \rangle$ 
and $|\Theta_2 \rangle$.
In the following we will use anti de Sitter coordinate
system (\ref{ads_coord_system}), in which these boundary states
correspond to the D-branes located at  $\psi=\psi_1$ and
$\psi=\psi_2$, respectively.
One way to compute the overlap is to insert the resolution of unity
(\ref{kre}).
This gives
\beq
  \langle \Theta_1 | \Theta_2 \rangle =
   \int \frac{d^2y}{(y-\bar y)^2}\int_0^\infty d\lambda\, A^2 \sin(i \Theta_1 \lambda)
             \sin(i \Theta_2 \lambda)=
  \frac{A^2 \pi}{2} \,\delta(\psi_1-\psi_2)
                     \int \frac{d^2y}{(y-\bar y)^2}.
\eeq
The same matrix element may be computed by using the coordinate
representation of the boundary wavefunction
\beq
  \langle g |\Theta_i \rangle =- A \pi \frac{\delta(\psi-\psi_i)}{\cosh \psi_i}.
\eeq
The SL(2,R) invariant measure on $H_3^+$ is simply a
volume form $[dg]=\cosh^2 \psi \cosh w \,d\psi \,d\omega \,dt$, so the overlap 
becomes
\beq
  \langle \Theta_1 | \Theta_2 \rangle =2 \pi \,  \frac{A^2 \pi}{2}
    \delta(\psi_1-\psi_2)
              \int dt \,dw\, \cosh w.
\eeq
Hence we conclude that  the divergent factors that appear in (\ref{dbiac_ads2})
and (\ref{sps_ads2}) match each other, up to a numerical factor,
and do not introduce extra dependence on the location of the D-brane.
Therefore the consistency of  (\ref{dbiac_ads2}) and (\ref{sps_ads2})
is not an artifact of the choice of a coordinate system.

\section{Discussion}
In this paper we showed that the conformal bootstrap 
on the disk \cite{GKS} gives boundary states that appear as 
surfaces of $X^i=const$ ($i=0,...,3$) in $H_3^+$.
The gluing conditions determine the coordinate $X^i$ that is
normal to the D-brane worldvolume, while the complex parameter
$\Theta$ that appears in the one-point function (\ref{fsin})
determines the value of $X^i$ where the D-brane is located.
In $H_3^+$, the two-sphere at $X^0=const$ corresponds to imaginary
$\Theta$, which is related via  (\ref{idtheta}) to
the continuous representation of SL(2,R).
The hyperbolic plane at $X^a=const$ ($a=1,2,3$) corresponds to
the $\Theta$ of the form (\ref{strangeresult}).
This value of $\Theta$ is consistent with the fusion of
the degenerate operator $\Phi_{-{1 \over 2}}$ to the boundary of the
worldsheet and spacetime.
The CFT partition sum on the disk for D-branes that we have studied
reproduces the DBI action up to a finite normalization constant.
Both quantities are infinite for D-branes that appear as hyperbolic
planes, reflecting the infinite volume of the latter.
However this volume infinity is shown to appear as a divergent integral
over the $y$-plane in the CFT description, and therefore may be isolated
and treated with the help of any suitable regulator.
\paragraph{}
A number of interesting things happen when one of the spacelike 
coordinates in $H_3^+$ is Wick rotated, giving Lorenzian $AdS_3$.
Two-spheres at $X^0=const$ turn into $dS_2$ surfaces.
The $X^0$ coordinate becomes unrestricted, so one may now have
$dS_2$ branes located at negative $X^0$.
Of course, the D-brane at  $X^0={\tilde C}$ and the one at  $X^0=-{\tilde C}$ 
are physically equivalent.
In the worldsheet description the manifestation of this equivalence
is that the one-point functions of these D-branes differ by a 
factor of $(-)^{2 h}$, which does not affect conformal bootstrap.
In $AdS_3$ it becomes possible to have D-branes located at $X^0<1$
($H_2$ instantonic branes).
These correspond to real $\Theta$, which is related via (\ref{idtheta})
to the principal discrete representation of SL(2,R).
Finally, D-branes at real $\X^3=const$ go into $AdS_2$ branes
under the Wick rotation of $X^1$ or $X^2$.
\paragraph{}
An interesting question  is the
interpretation of the parameter $\Theta$ which labels $AdS_2$
branes, and is given by (\ref{strangeresult}).
One possibility may be that in the open string channel the expression
(\ref{mb2}) represents  the sum of
the principal continuous character and its spectral flow
by one unit, but the precise identification needs more work\footnote{The
precise interpretation of (\ref{mb2}) in the open string
channel is difficult, since the (spectral flowed) character of
the principal continuous representation is divergent.}.
Another important question that would be interesting
to analyze is the geometric interpretation of 
boundary states that correspond to degenerate
representations of SL(2,R).
It is clear that corresponding D-branes must be localized in $AdS_3$, since
the spectrum of open strings living on them contains a finite
number of current algebra blocks.
The possibility that they appear as two-spheres in $H_3^+$,
mimicking the SU(2) case, seems to be ruled out by the identification
of such two-spheres with extended D-branes that correspond to
principal continuous series.
Note that up to the normalization factor, which depends only on the
half-integer $h'$ that labels the boundary state,
the one-point functions for the boundary states that correspond
to the degenerate and principal discrete
representations of SL(2,R) are equal.
The semiclassical results for boundary wavefunctions are therefore 
indistinguishable.
The meaning of this is not clear.
\paragraph{}
Let us make a brief comment about the role of the spectral flow
in constructing D-branes in the $H_3^+$ WZNW model.
Recall, that the ratio of structure constants (\ref{sc_rel}).
has in its right-hand side the square root
of the ratio of two-point functions $\langle \Phi_h \Phi_h \rangle$
and $\langle \Phi_{{k \over 2}-h} \Phi_{{k \over 2}-h} \rangle$ 
[see also (\ref{structur})]
This ratio is essentially making up for the relative 
normalization of the operators $\Phi_h$ and
$ \Phi_{{k \over 2}-h}$.
This leads us to the observation, that up to a phase,
\beq
\label{urel}
  \frac{U({k \over 2}-h)}{
   \sqrt{\langle \Phi_{{k \over 2}-h} \Phi_{{k \over 2}-h} \rangle}}
 \sim
  \frac{U(h)}{
   \sqrt{\langle \Phi_h \Phi_h \rangle}}.
\eeq
This implies that D-branes that correspond to degenerate representations
of SL(2,R) are similar to D-branes in the SU(2) WZNW model.
The relation (\ref{urel}) also hints that  D-branes 
in the SL(2,R)/U(1) parafermion theory may be
quite similar to their SU(2)/U(1) counterparts\footnote{Recently  D-branes in the coset model were studied from
the geometric point of view \cite{ES}.}.
Understanding  D-branes in SL(2,R)/U(1) is, of course, a
very interesting open problem.

\section*{Acknowledgements}
We gratefully acknowledge discussions with D. Kutasov.
We would also like to thank D. Kutasov and A. Schwimmer for
comments on the manuscript.
This work was supported, in part, by DOE grant \#DE-FG02-90ER40560.

\appendix{The solutions of the Knizhnik-Zamolodchikov equation}
In this appendix we write down the solutions of the Knizhnik-Zamolodchikov
equation (\ref{kzeq}), which we reproduce below
\begin{eqnarray}
\label{kzeq_a}
&&\Bigg[ -z (z-1) (k-2) \partial_z
    +y (y-1) (z-y) \partial_y^2 \\ \nonumber
&&   \quad  +\left[ (\Delta{+}1) (-y^2{+}2 z y{-}z){-}2 h_0 y (y{-}1)
           {-}2 y h_1 (z{-}1) {-}2 (y{-}1) h_2 z \right] \partial_y
          \\ \nonumber
&&   \quad  {+}\left[2 h_0 \Delta (z{-}y) - 2 h_0 h_1 (z{-}1){-}2 h_0 h_2 z \right]
\Bigg]  H\left(\bega{ll} h_0 &h_1\\ h_2 &h_3\ea,y,z\right)=0
\end{eqnarray}
for the special cases that appear in the main text.
We first consider the conformal blocks that appear in the
two-point function, which can be written in the $s$ and $t$ channels as
\begin{equation}
\label{2ptfun_as}
 \langle \Phi_{-\frac{1}{2}}(y_1,z_1) \Phi_{h}(y_2,z_2)  \rangle=
\frac{|y_2-{\bar y_2}|^{-1-2h}}
                            {|y_1-{\bar y_2}|^{-2}} \;
                    \frac{|z_2-{\bar z_2}|^{-\frac{3 u}{2}-2\Delta_h}}
                            {|z_1-{\bar z_2}|^{-3 u}}
      \;  H\left(\bega{ll} -1/2 &h\\ -1/2 &h\ea,y,z\right)  
\end{equation}
and
\begin{equation}
\label{2ptfun_at}
 \langle \Phi_{-\frac{1}{2}}(y_1,z_1) \Phi_{h}(y_2,z_2)  \rangle=
\frac{|y_2-{\bar y_2}|^{-1-2h}}
                            {|y_1-{\bar y_2}|^{-2}} \;
                    \frac{|z_2-{\bar z_2}|^{-\frac{3 u}{2}-2\Delta_h}}
                            {|z_1-{\bar z_2}|^{-3 u}}
      \;  H\left(\bega{ll} {-}1/2 & {-}1/2\\ h &h\ea,1{-}y,1{-}z\right)  
\end{equation}
respectively, where $y$ and $z$ are the cross-ratios defined in the
main text.
The conformal blocks of this sort have been previously known
\cite{T,GK,FZ}, so we present the results here for completeness.
\paragraph{}
Consider $h_0=-{1 \over 2}$. It is not hard to see that 
\beq
\label{anz_h}
H\left(\bega{ll} -1/2 &h_1\\ h_2 &h_3\ea,y,z\right)  = 
         C_1[y H_1^+(z)+H_0^+(z)]+ C_2[y H_1^-(z)+H_0^-(z)],
\end{equation}
where $C_1$ and $C_2$ are some constants.
One can convince oneself that this is the case by substituting 
the expression (\ref{anz_h}) into (\ref{kzeq_a}) and noting
that the ${\cal O}(y^2)$ term is trivially zero. 
The vanishing of  ${\cal O}(y)$ and  ${\cal O}(1)$ terms implies that
$H_1^\pm(z)$ and  $H_0^\pm(z)$ satisfy 
\begin{eqnarray}
\label{heq1}
&& \left[ -z(z-1) (k-2) \partial_z+
      \left(\left(\frac{3}{2}-h_3\right) z+h_1-1\right) \right] H_1^\pm 
   +\Delta H_0^\pm=0, \\
\label{heq2}
 &&-z (z{-}1) (k{-}2) \partial_z H_0(z)
                   +\left[h_2{-}h_1{+}h_3{-}\frac{1}{2} \right] H_1^\pm
   +\left[ \left(h_3{+}\frac{1}{2}\right) z {-}h_1\right] H_0^\pm=0.
\end{eqnarray}
Substituting the first equation into the second
one obtains the  second-order differential equation for $H_1^\pm(z)$,
which has two linearly independent solutions.
(This explains the need for the superscript $\pm$.)
After the substitution
\begin{equation}
  H_1^\pm(z)=z^{\beta_1^\pm} (1-z)^{\beta_2^\pm} K^\pm(z),
\label{subh}
\end{equation}
\begin{equation}
\label{betas}
 \beta^{+}_i=u(1-h_i), \qquad \beta^{-}_i=u h_i.
\end{equation}
One has the hypergeometric equation for  $K^\pm(z)$ 
\begin{equation}
 \left[  z(1-z) \partial_z^2+(C-(A+B+1) z)\partial_z-A B \right] K^\pm(z)=0,
\label{hgeq}
\end{equation}
where 
\begin{eqnarray}
\label{abc}
A+B+1&=&2\left( \beta_1+\beta_2+\frac{k-3}{k-2} \right), \\ \nonumber
A B&=&\left(\beta_1+\beta_2\right) \left(\beta_1+\beta_2-1+
                           2 \frac{k-3}{k-2}\right)
          +\frac{ \left( \frac{3}{2}-h_3\right)}{k-2} \left(
          \frac{h_3+\frac{1}{2}}{k-2}-1\right), \\ \nonumber
C&=&2 \beta_1+\frac{k-3}{k-2}.
\end{eqnarray}
The solution of (\ref{hgeq}) that is regular at $z=0$
is a hypergeometric function $F(A,B,C,z)$ with $A$, $B$, $C$ given
by the solution of (\ref{abc}).
Using (\ref{heq1}), (\ref{subh}), (\ref{betas}) and (\ref{abc}),
together with the identification $h_1=h_3=h$, $h_2=-1/2$, $\Delta=-1$,
one may reproduce the conformal blocks (\ref{cblock1})-(\ref{cblock4}).
The alternative identification $h_2=h_3=h$, $h_1=-1/2$, 
gives rise to
\beq
H\left(\bega{ll} {-}1/2 & {-}1/2\\ h &h\ea,1{-}y,1{-}z\right) {=}
   {\tilde C_1}
      (1-z)^{\frac{3 u}{2}} z^{u(1-h)} F(u,1{+}2 u(1{-}h),1{+}2u,1{-}z) 
        (1{-}y){+}{\cdots},
\eeq
where ${\tilde C_1}$ is some constant, and the dots stand for 
conformal blocks whose explicit form will not be needed to us.
\paragraph{} 
The equation (\ref{kzeq_a}) can also be solved for the case
$h_0=-\frac{k+2}{2}$ \cite{T,GK}.
The solution involves hypergeometric functions of two arguments.
We will instead look at the conformal block with  $h_3=\frac{k+1}{2}$.
It was shown in \cite{FZ} that the relevant solution
takes a simple form:
\begin{eqnarray}
\label{anz_h2}
  H\left(\bega{ll} h_0 &h_1\\ h_2 &(k+1)/2\ea,y,z\right)&=&
  C_1'[(y-z)^{-\Delta} H_0'^+(z)+(y-z)^{-\Delta-1} H_1'^+(z)] \\ \nonumber
&& \qquad  +C_2'[(y-z)^{-\Delta} H_0'^-(z)+(y-z)^{-\Delta-1} H_1'^-(z)].
\end{eqnarray}
Again, this can be easily seen by substituting (\ref{anz_h2}) 
into (\ref{kzeq_a}) and observing that the ${\cal O}(-\Delta+1)$ 
and ${\cal O}(-\Delta-2)$  terms vanish trivially. 
The requirement for the vanishing of ${\cal O}(-\Delta)$ and ${\cal O}(-\Delta-1)$ 
terms leads to the system of two differential equations
similar to (\ref{heq1})--(\ref{heq2}):
\begin{eqnarray}
\label{heq1_dual}
 &&  \left[ z(z-1)(k-2) \partial_z + a
               \right] H_0'^\pm(z) +\left[ \Delta+1-2 h_0 \right]  H_1'^\pm(z)=0,\\
\label{heq2_dual}
 &&  \left[ z (z-1) (k-2) \partial_z +b \right] H_1'^\pm(z)-\Delta z (z-1) H_0'^\pm(z)=0, 
\end{eqnarray}
where
\begin{eqnarray}
 a&=&z \left[ \Delta (2 h_2+1-k-2 \Delta)+2 h_0 (h_1+h_2) \right]+
           \Delta (\Delta-2 h_2+k)-2 h_0 h_1, \\
 b&{=}&z \left[ (\Delta{+}1) (3{-}2 h_0{-}k){+}2 h_0 (h_1{+}h_2) \right]-
  (\Delta{+}1) (2 h_2{+}1{-}k{-}\Delta){-}2 h_0 h_1.
\end{eqnarray}   
Consider  the two-point function dual to (\ref{2ptfun_as})
\begin{eqnarray}
\label{2corr_2_a}
\langle \Phi_{h}(y_1,z_1)  \Phi_{\frac{k+1}{2}}(y_2,z_2)  \rangle&=&
|y_2{-}{\bar y_2}|^{-k-1+2h}  |y_1{-}{\bar y_2}|^{-4 h} \\ \nonumber && \; \;
                    |z_2{-}{\bar z_2}|^{2\Delta_h-2\Delta_{\frac{k+1}{2}} }
                            |z_1{-}{\bar z_2}|^{-4 \Delta_h} 
       H\left(\bega{ll} h &(k+1)/2\\ h &(k+1)/2\ea,y,z\right).  
\end{eqnarray}
The analogs of (\ref{subh})-(\ref{betas}) 
for the conformal blocks that appear in the expansion (\ref{anz_h2})  of
$ H\left(\bega{ll} h &(k+1)/2\\ h &(k+1)/2\ea,y,z\right)$ 
are
\begin{eqnarray}
 && H_1'^\pm(z)=z^{\beta_1^\pm} (1-z)^{\beta_2^\pm} K'^\pm(z),  \\
 &&\beta_1^+=u h (k-1), \qquad \beta_2^+= 2  u h (h-1), \\ \nonumber
 &&\beta_1^-=1+h+u(1-h), \qquad \beta_2^-= 2 u h (h-1).
\end{eqnarray}
Solving (\ref{heq1_dual})--(\ref{heq2_dual}) gives
(\ref{cblock1_dual})--(\ref{cblock4_dual}).
We will also need the conformal block that appears in the boundary
expansion of the two-point function (\ref{2corr_2_a})
[This expression is dual to (\ref{2ptfun_at}) ]
\begin{eqnarray}
\label{2corr_2_at}
&&\langle \Phi_{h}(y_1,z_1)  \Phi_{\frac{k+1}{2}}(y_2,z_2)  \rangle=
|y_2{-}{\bar y_2}|^{-k-1+2h}  |y_1{-}{\bar y_2}|^{-4 h} \\ \nonumber && \; \;
  \qquad   \qquad    |z_2{-}{\bar z_2}|^{2\Delta_h-2\Delta_{\frac{k+1}{2}} }
                            |z_1{-}{\bar z_2}|^{-4 \Delta_h} 
       H\left(\bega{ll} h & h\\ (k+1)/2 &(k+1)/2\ea,1-y,1-z\right).  
\end{eqnarray}
The relevant solution is
\begin{eqnarray}
&&  H\left(\bega{ll} h & h\\ (k+1)/2 &(k+1)/2\ea,1-y,1-z\right)= \\ \nonumber
   && \qquad \qquad {\tilde C}'_1
\left( [1{-}x]{-}[1{-}z)] \right)^{-2h} (1{-}z)^{2 u h (h-1)} z^{h+uh}
                               F(u,2 u h, 1{+}2 u, 1{-}z)
+\cdots.
\end{eqnarray}


\appendix{Spectral flow and Parafermionic Construction}

In this appendix we show how three and four point correlation functions
of the primaries of the $SL(2,R)$ WZNW model transform under the spectral 
flow by one unit.
As it was shown in \cite{MO} the
spectral flow by one unit ($w=1$) maps the highest weight representation into
lowest weight representation and lowest weight($w=-1$) into highest weight:
\beq
\bega{l}
\Phi^{hw}_{h,-h}\rightarrow\Phi^{lw}_{{k\over 2}-h,{k\over 2}-h},\\
\Phi^{lw}_{h,h}\rightarrow\Phi^{hw}_{{k\over 2}-h,-{k\over 2}+h}.
\ea
\eeq
For simplicity we are suppressing the antiholomorphic part of the operators.
The operators of the current algebra corresponding to the highest weight
representation can be expressed in terms of  free bosonic field (with the wrong sign
in the propagator in the contrast to the $SU(2)$ case) and so called parafermionic
operators.
\beq\label{para}
\Phi^{hw}_{h; m}=e^{i m\sqrt{2\over k}\phi(z)}\Psi^{hw}_{h;m};\quad m\in
-h+Z. \eeq
where $\Phi^{hw}_{h; m}$ are defined as follows
\beq\label{paraf}
\bega{l}
\Phi^{hw}_{h; m}(z)\equiv (J^-_0)^{-h-m}\Phi^{hw}_{h; -h}(z)\quad m<-h,\\
\Phi^{hw}_{h; m}(z)\equiv (J^+_{-1})^{m+h}\Phi^{hw}_{h; -h}(z)\quad
m\geq-h. \ea
\eeq
Note that the highest weight field
$\Phi^{hw}(y,z)$ can be written in terms of $\Phi^{hw}_{h;m}(z)$
in the following way
\beq
\Phi^{hw}(y,z)=\sum_{m=-\infty}^{-h}\Phi^{hw}_{h;m}y^{-h-m}(z).
\eeq
Now using (\ref{paraf}) one can easily derive the conformal weights
for parafermions
\beq\label{dim}
\Delta_{\Psi^{hw}_{h,m}}=\left\{\bega{l} -{h(h-1)\over k-2}+{m^2\over
k};\quad m\leq-h\\ -{h(h-1)\over k-2}+{m^2\over k}+(m+h);\quad m\geq -h
\ea\right\}.
\eeq
Similarly for the lowest weight representation we have the lowest weight
parafermions which have the following conformal dimensions
\beq\label{dim1}
\Delta_{\Psi^{lw}_{h,m}}=
\left\{\bega{l} -{h(h-1)\over k-2}+{m^2\over k};\quad m\geq h\\
-{h(h-1)\over k-2}+{m^2\over k}+(h-m);\quad m\leq h
\ea\right\}.
\eeq
The parafermionic theory is believed to be unitary (see for example
\cite{DLP}). From the expressions (\ref{dim}), (\ref{dim1}) we see that the
operators $\Psi^{lw}_{{k\over 2};{k\over 2}}$ and $\Psi^{hw}_{{k\over
2};-{k\over 2}}$ have zero conformal weight and should be proportional to the
identity operator. It is also natural to assume that $\Psi^{hw}_{h;m}(z)$ and
$\Psi^{lw}_{{k\over 2}-h;m+{k\over 2}}(z)$ are identified up to a multiplicative
constant (which in general depends on $(h,m)$ and is related to the normalization of the operators)
since they have the same worldsheet dimension. We will need in particular 
relation between $\Psi^{hw}_{h;-h}(z)$ and its counterpart under the spectral flow
\beq
\Psi^{hw}_{h;-h}(z)=
B(h)\Psi^{lw}_{{k\over 2}-h;{k\over 2}-h}(z).
\eeq
The relative normalization $B(h)$ of parafermions adopted in this paper is
\beq
B(h)=\sqrt{\langle \Phi_h \Phi_h \rangle\over
\langle \Phi_{{k \over 2}-h} \Phi_{{k \over 2}-h} \rangle},
\eeq
where ${\langle \Phi_h \Phi_h \rangle}$ is the two point function of the operator $\Phi_h$ stripped of
the worldsheet and spacetime coordinates dependence. 
\paragraph{}
This assumption will allow us to derive the transformation 
properties of three and four point functions under spectral flow.
One can use the conformal symmetry (spacetime and worldsheet) to write a 
three point correlation function in the following form
\beq\label{three}
\langle\Phi_{h_1}(y_1,z_1)\Phi_{h_2}(y_2,z_2)\Phi_{h_3}(y_3,z_3)\rangle=
D(h_1,h_2,h_3)\prod_{i<j}^3 y^{\lambda_{ij}}z^{\Delta_{ij}},
\eeq
where $y_{ij}=y_i-y_j$ and
\begin{eqnarray}
\lambda_{12}=h_1+h_2-h_3,\\
\lambda_{13}=h_1+h_3-h_2,\\
\lambda_{23}=h_2+h_3-h_1.
\end{eqnarray}
$z_{ij}$ and $\Delta_{ij}$ are defined in the similar 
manner with $h_i\rightarrow\Delta_i$.
The coefficient $D(h_1,h_2,h_3)$ is symmetric under permutation of its arguments 
since correlation function should not depend on the order of operators and    
can be easily extracted from (\ref{three})
\beq
D(h_1,h_2,h_3)=\lim_{\bega{l}y_3\rightarrow \infty\\ z_3\rightarrow \infty\ea}
\langle\Phi_{h_1}(0,0)\Phi_{h_2}(1,1)\Phi_{h_3}(y_3,z_3)\rangle y_3^{2h_3}z_3^{2\Delta_3}.
\eeq
So we conclude that in order to compute $D(h_1,h_2,h_3)$ one needs to know correlators 
of the form
\beq
\langle\Phi_{h_1}(0,0)|\Phi_{h_2}(1,1)|\Phi_{h_3}(\infty,\infty)\rangle=
\langle\Phi_{h_1;-h_1}(0)|\Phi_{h_2}(1,1)|\Phi_{h_3;h_3}(\infty)\rangle.
\eeq
Expanding the r.h.s. of the above equation in spacetime coordinate 
and using parafermion representation we will get
$$
\langle\Phi_{h_1;-h_1}(0)|\Phi_{h_2}(1,1)|\Phi_{h_3;h_3}(\infty)\rangle=
\sum_{m_2}\langle\Phi_{h_1;-h_1}(0)|\Phi_{h_2;m_2}(1)|\Phi_{h_3;h_3}(\infty)\rangle=
$$
$$
\sum_{m_2}\langle\Psi_{h_1;-h_1}(0)|\Psi_{h_2;m_2}(1)|\Psi_{h_3;h_3}(\infty)\rangle
\delta(m_2+h_3-h_1)=
$$
\beq\label{specthree}
\sum_{m_2}B(h_1)B(h_3)
\langle\Psi_{{k\over 2}-h_1;{k\over 2}-h_1}(0)|\Psi_{h_2;m_2}(1)
|\Psi_{{k\over 2}-h_3;-{k\over 2}+h_3}(\infty)\rangle
\delta(m_2+(-{k\over 2}+h_3)+({k\over 2}-h_1))=
\eeq
$$
B(h_1)B(h_3)\langle\Phi_{{k\over 2}-h_1;{k\over 2}-h_1}(0)|\Phi_{h_2}(1,1)|
\Phi_{{k\over 2}-h_3;-{k\over 2}+h_3}(\infty)\rangle.
$$
In the derivation above we used the free field three point function
which is just a delta function.
One can convince oneself that the last line of (\ref{specthree})
is equal to $B(h_1)B(h_3) D(k/2-h_1,h_2,k/2-h_3)$, so we obtained the following relation
\beq\label{threeres}
D(h_1,h_2,h_3)=B(h_1)B(h_3)D(k/2-h_1,h_2,k/2-h_3).
\eeq
Using the explicit expression for $D(h_1,h_2,h_3)$ from \cite{T} one can check that (\ref{threeres})
indeed holds.


Let us now find the transformation of
four point correlation function under the spectral flow.
For the rest of this appendix we will set $B(h)$ to one to simplify
formulae (one can easily restore the overall normalizations
guided by the example of three point function).
Recall that using the conformal symmetry
(both worldsheet and spacetime)(cr.
(\ref{ward})) the four point function can be written as
\beq\label{conf}
\langle\Phi_{h_1}(y_1,z_1)\Phi_{h_0}(y_0,z_0)\Phi_{h_2}(y_2,z_2)\Phi_{h_3}(y_3,z_3)\rangle=
\prod y_{ij}^{\mu_{ij}}z_{ij}^{\nu_{ij}}
H\left(\bega{ll} h_0 &h_1\\ h_2 &h_3\ea,y,z\right),
\eeq
where $H$ is defined as
\beq\label{conf1}
H\left(\bega{ll} h_0 &h_1\\ h_2 &h_3\ea,y,z\right)=
\lim_{\bega{l}y_3\rightarrow \infty\\ z_3\rightarrow \infty\ea}
\langle\Phi_{h_1}(0,0)\Phi_{h_0}(y,z)\Phi_{h_2}(1,1)\Phi_{h_3}(y_3,z_3)\rangle
y_3^{2h_3} z_3^{\Delta_3}.
\eeq
We see that in order to compute $H$ one needs to know correlators of the
form
\beq
\langle\Phi_{h_1}(0,0)|\Phi_{h_0}(y,z)\Phi_{h_2}(1,1)|\Phi_{h_3}(\infty,\infty)\rangle=
\langle\Phi_{h_1;-h_1}(0)|\Phi_{h_0}(y,z)\Phi_{h_2}(1,1)|\Phi_{h_3;h_3}(\infty)\rangle.
\eeq
We will also need the following relation
$$
\langle\Phi_{h_1}(\infty,0)|\Phi_{h_0}(y,z)\Phi_{h_2}(1,1)|\Phi_{h_3}(0,\infty)\rangle=
y^{-2h_0}H\left(\bega{ll} h_0 &h_1\\ h_2 &h_3\ea,{1\over y},z\right)=
$$
\beq\label{rev}
\langle\Phi_{h_1;h_1}(0)|\Phi_{h_0}(y,z)\Phi_{h_2}(1,1)|\Phi_{h_3;-h_3}(\infty)\rangle,
\eeq
To prove (\ref{rev}) let us rewrite (\ref{conf}) with
$y_1\leftrightarrow y_3$
\beq\label{conf2}
\langle\Phi_{h_1}(y_3,z_1)\Phi_{h_0}(y_0,z_0)\Phi_{h_2}(y_2,z_2)\Phi_{h_3}(y_1,z_3)\rangle=
\prod y_{ij}^{\tilde\mu_{ij}}\prod z_{ij}^{\nu_{ij}}
H\left(\bega{ll} h_0 &h_1\\ h_2 &h_3\ea,\tilde y,z\right),
\eeq
where nonzero $\tilde\mu_{ij}$
\begin{eqnarray}
\tilde\mu_{01}&=&-2 h_0, \\
\tilde\mu_{13}&=&-h_1-h_3+h_0+h_2, \\
\tilde\mu_{12}&=&-h_2-h_3+h_0+h_1, \\
\tilde\mu_{23}&=&-h_0-h_1-h_2+h_3,
\end{eqnarray}
are obtained from $\mu_{ij}$
simply by exchange of indices $1\leftrightarrow 3$ and
$\tilde y={1\over y}$. Taking the proper limit in (\ref{conf2})
we arrive to the expression (\ref{rev}).

Now, using (\ref{paraf}) and expanding the
operators in spacetime coordinate, we have
$$
\sum_{m_0,m_2}
\langle\Phi_{h_1;-h_1}(0)|\Phi_{h_0;m_0}(z)\Phi_{h_2;m_2}(1)|\Phi_{h_3;h_3}(\infty)\rangle y^{-h_0-m_0}=
$$
$$
\sum_{m_0,m_2}
\langle\Psi_{h_1;-h_1}(0)|\Psi_{h_0;m_0}(z)\Psi_{h_2;m_2}(1)|\Psi_{h_3;h_3}(\infty)\rangle y^{-h_0-m_0}\times
$$
\beq
\langle
e^{i\sqrt{2\over k}(-h_1)\phi(0)}|e^{i\sqrt{2\over
k}(m_0)\phi(z)}e^{i\sqrt{2\over k}(m_2)\phi(1)}|e^{i\sqrt{2\over
k}(h_3)\phi(\infty)}\rangle=
\eeq
$$
\sum_{m_0,m_2}
\langle\Psi_{h_1;-h_1}(0)|\Psi_{h_0;m_0}(z)\Psi_{h_2;m_2}(1)|\Psi_{h_3;h_3}(\infty)\rangle y^{-h_0-m_0}
z^{m_0h_1{2\over k}}(1-z)^{-{2\over k}m_0m_2}.
$$
In the second line we used the representation of the operators of
$SL(2,R)$ in terms of free field and parafermions, and in
the last line the four point function for
operators in the free field theory was used.

After performing the spectral flow to the ``in'' and ``out'' states we will
obtain
$$
\langle\Phi_{{k\over 2}-h_1}(\infty, 0)|
\Phi_{h_0}(y,z)\Phi_{h_2}(1,1)
|\Phi_{{k\over 2}-h_3}(0,\infty)\rangle=
$$
$$
\sum_{m_0,m_2}
\langle\Phi_{{k\over 2}-h_1;{k\over 2}-h_1}(0)|
\Phi_{h_0;m_0}(z)\Phi_{h_2;m_2}(1)
|\Phi_{{k\over 2}-h_3;-{k\over 2}+h_3}(\infty)\rangle
y^{-h_0-m_0}=
$$
\beq\label{specsym}
\sum_{m_0,m_2}
\langle\Psi_{h_1;-h_1}(0)|\Psi_{h_0;m_0}(z)\Psi_{h_2;m_2}(1)|\Psi_{h_3;h_3}(\infty)\rangle y^{-h_0-m_0}
z^{-{2\over k}({k\over 2}-h_1)m_0}(1-z)^{-{2\over k}m_0m_2}=
\eeq
$$
\langle\Phi_{h_1}(0,0)|\Phi_{h_0}(yz,z)\Phi_{h_
2}(1,1)|\Phi_{h_3}(\infty,\infty)\rangle z^{h_0}.
$$
Using (\ref{rev}), (\ref{specsym}) can be rewritten in the
following form:
\beq\label{specsym1}
y^{h_0}H\left(\bega{ll} h_0 &{k\over 2}-h_1\\ h_2 &{k\over
2}-h_3\ea,y,z\right)=
\left(z\over y\right)^{h_0}H\left(\bega{ll} h_0 &h_1\\ h_2
&h_3\ea,{z\over y},z\right). \eeq
In order to check the relation (\ref{specsym1}) we recall that in the Appendix A 
(see also (\ref{cblock1_dual})-(\ref{cblock4_dual})) we computed
\beq 
H\left(\bega{ll} -1/2 &h\\ -1/2 &h\ea,y,z\right)=H_0(z)+yH_1(z),
\eeq
\beq
H\left(\bega{ll}  h &k/2+1/2\\ h &k/2+1/2\ea,y,z\right)=(y-z)^{-2h}H_0^\prime(z)+
(y-z)^{-2h-1}H_1^\prime(z).
\eeq
In order to compare these two expressions we should make some
transformations. By changing the positions of operators in (\ref{conf}) and
demanding that the result do not change under this transformation we have
\beq\label{half}
H\left(\bega{ll} h&-1/2\\
h&-1/2\ea,y,z\right)=
(1-y)^{-2h-1}(1-z)^{-2\Delta_h+2\Delta_{-1/2}}
H\left(\bega{ll} -1/2 &h\\ -1/2 &h\ea,y,z\right).
\eeq
Substituting (\ref{half}) into (\ref{specsym1}) we will get
\beq\label{sol}
\bega{l}
H_0^\prime(z)=z^h(1-z)^{-2\Delta_h+2\Delta_{-1/2}}H_0(z),\\
H_1^\prime(z)=z^{h+1}(1-z)^{-2\Delta_h+2\Delta_{-1/2}}(H_0(z)+H_1(z)).
\ea 
\eeq
A straightforward comparison of (\ref{cblock2}), (\ref{cblock4}) with
(\ref{cblock1_dual}), (\ref{cblock3_dual}) verifies the first line in the above equation.
The verification of the second line is technically more involved and 
one needs to employ Gauss' recursion formulas for hypergeometric functions.
Using (\ref{gauss_1}), (\ref{cblock3}), (\ref{cblock4}), (\ref{cblock4_dual})
 we see that
\beq
{H_1^\prime}^-(z)=z^{h+1}(1-z)^{-2\Delta_h+2\Delta_{-1/2}}(H_0^-(z)+H_1^-(z))
\eeq
holds. Finally we can combine (\ref{gauss_4}),(\ref{cblock1}),(\ref{cblock2}),(\ref{cblock2_dual}) 
to show that
\beq
{H_1^\prime}^+(z)=z^{h+1}(1-z)^{-2\Delta_h+2\Delta_{-1/2}}(H_0^+(z)+H_1^+(z)).
\eeq
This concludes our check.

\appendix{Some useful integrals}
\paragraph{}
In this Appendix we compute some integrals that were important
for the discussion in section 3.
The integrals of this sort were frequently encountered
in section 4, which we will heavily borrow from.
We start by computing the integral that appears in (\ref{refsymm3}).
\beq
\int d^2 y' \, \frac{|y-y'|^{-4 h}}{(1+y' \bar y')^{2-2 h}}=
  \int d^2 y' \left({y'_1}^2+{y'_2}^2-2 R y'_1+R^2 \right)^{-2 h}
       \left( 1+{y'_1}^2+{y'_2}^2 \right)^{2 h-2},
\eeq
where we introduced $R=|y|$. 
Exponentiating the integrand, completing the square, and
performing the integral over the $y'$-plane one obtains
\beq
  \frac{\pi}{\Gamma(2 h) \Gamma(2-2 h)} \int_0^\infty \, 
  \frac{ t^{2 h-1} s^{1-2 h}}{t+s} \,
  \exp \left( \frac{R^2 t^2}{t+s}-t R^2-s \right).
\eeq
Writing $s= \alpha t$, as usual, and integrating over $t$ leaves
us with
\beq
 \frac{\pi}{\Gamma(2 h) \Gamma(2-2 h)} \int_0^\infty d \alpha
   \frac{ \alpha^{-2 h}}{\alpha+(1+R^2)},
\eeq
which depends only on $(1+y \bar y)$, as it should.
After rescaling the integral above becomes a simple Veneziano
integral which is not hard to do.
The result is
\beq
 \frac{\pi}{1-2 h} \frac{1}{(1+y \bar y)^{2 h}}.
\eeq
Substituting this into (\ref{refsymm3}) we recover 
the advertised result
\beq
  U(h)=-{\cal R}(h) U(1-h).
\eeq
This result is important for fixing the solution of
the bootstrap equation to have the form (\ref{fsin})
\paragraph{}
Let us now compute the $h$-dependent factor that appears in the 
analog of (\ref{mb}) for the partition sum of an open string
stretched between a D-brane described by the one-point
function of the form 
$\langle \Phi_h(y ,\bar y) \rangle =\frac{U(h)}{y-\bar y)^{2h}}$
and the basic D-instanton.
This factor is given by the following integral
\beq
\label{modmb}
\int d^2 y \, (y- \bar y)^{2-2 h} \,(1+y \bar y)^{-2 h}.
\eeq
In fact, this integral may be computed in two ways.
First let us notice that the expression above can be represented
by an integral of the form (\ref{someint}) with $X^0=X^1=X^2=0$
and $X^3=1$.
In the parameterization used in  section 4 this 
corresponds to ${\tilde \psi}=i {\pi \over 2}$.
The value of this integral can be read from (\ref{iii1})
\beq
\label{c7}
   \frac{\pi}{2 h-1} \, \frac{\sinh[(2 h-1) {\tilde \psi}]}{\sinh {\tilde \psi}}=
           \frac{\pi}{2 h-1} \, \sinh[i (2 h-1) \frac{\pi}{2}].
\eeq
The alternative computation uses the integral that appears in
(\ref{sam_ads2}) with the identification $X^0=1$, $X^1=X^2=X^3=0$.
This corresponds to $\psi=0$, and using (\ref{iii2})
we recover (\ref{c7}).
Substituting (\ref{c7}) into (\ref{mb}) gives (\ref{mb2}).

\appendix{Some properties of hypergeometric functions}
In this appendix we give some formulas that are used in this paper.
We will need the following Gauss recursion formulas for hypergeometric functions \cite{GR}
\begin{eqnarray}\label{gauss_1}
cF(a,b,c;z)+(b-c)F(a+1,b,c+1;z)-b(1-z)F(a+1,b+1,c+1;z)=0,\\
cF(a,b,c;z)-cF(a+1,b,c;z)+bzF(a+1,b+1,c+1;z)=0,\\
\label{gauss_2}
cF(a,b,c;z)-(c-b)F(a,b,c+1;z)-bF(a,b+1,c+1;z)=0.
\label{gauss_3}
\end{eqnarray}
In Appendix B we also needed the following relation that hypergeometric
functions satisfy
\beq\label{gauss_4}
F(a,b,b-a;z)+{a\over b-a}zF(a+1,b,b-a+1;z)-(1-z)F(a+1,b,b-a;z)=0,
\eeq
which we prove here.
\beq
\bega{l}
F(a,b,b-a;z)+{a\over b-a}zF(a+1,b,b-a+1;z)-(1-z)F(a+1,b,b-a;z)=\\
z\left[F(a+1,b,b-a;z)+{a\over b-a}F(a+1,b,b-a+1;z)-{b\over b-a}F(a+1,b+1,b-a+1;z)\right]=\\
{z\over b-a^\prime+1}[(b-a^\prime+1)F(a^\prime,b,b-a+1;z)-
bF(a^\prime,b+1,b-a+2;z)-\\
(b-a^\prime+1-b)F(a^\prime,b,b-a^\prime+2;z)]=0,
\ea
\eeq
where $a^\prime=a+1$. To get the second line we used (\ref{gauss_2})
and the last equality follows from (\ref{gauss_3}).

We are also giving the transformation properties of hypergeometric
functions.
Under $z\to 1/z$:
$$
F(a,b,c;z)=
{\Gamma(c)\Gamma(b-a)\over\Gamma(b)\Gamma(c-a)}
(-{1\over z})^a F(a,a+1-c;a+1-b;{1\over z})
+
$$
\beq
{\Gamma(c)\Gamma(a-b)\over\Gamma(a)\Gamma(c-b)}
(-{1\over z})^b F(b,b+1-c;b+1-a;{1\over z})~.
\eeq
Under $z\to 1-z$:
$$
F(a,b,c;1-z)=
{\Gamma(c)\Gamma(c-b-a)\over
\Gamma(c-a)\Gamma(c-b)}
F(a,b,a+b+1-c;z)
+$$
\beq
\label{hgzzz}
z^{c-a-b}
{\Gamma(c)\Gamma(a+b-c)\over
\Gamma(a)\Gamma(b)}
F(c-a,c-b,c+1-a-b;z)~.
\eeq


\end{document}